# Global Galerkin method for stability studies in incompressible CFD and other possible applications

## Alexander Gelfgat

**Abstract** In this paper the author reviews a version of the global Galerkin that was developed and applied in a series of earlier publications. The method is based on divergence-free basis functions satisfying all the linear and homogeneous boundary conditions. The functions are defined as linear superpositions of the Chebyshev polynomials of the first and second klinds that are combined into divergence free vectors. The description and explanations of treatment of boundary conditions inhomogeneities and singularities are given. Possible implementation for steady state solvers, path-continuation, stability solvers and straight-forward integration in time are discussed. The most important results obtained using the approach are briefly reviewed and possible future applications are discussed.

## 1    Introduction

This paper revisits a version of the global Galerkin method whose development started in the author's PhD thesis (Gelfgat, 1988) and resulted in an effective approach for stability analysis of model incompressible flows, so that well known and less known results had been

School of Mechanical Engineering,
Faculty of Engineering, Tel-Aviv
University, Tel-Aviv, Israel
e-mail: gelfgat@tau.ac.il

published continuously between 1994 and 2005. Several years ago I was asked to present this approach again, and it triggered some interest among several young colleagues. It is mainly their interest that inspired me to review all that was done and to make a consistent description of all the technical details, which were distributed over several papers, and sometimes omitted. Naturally, I have added some comments on my personal current opinion of what was done and what possibly can be done in the future.

The development of this approach started in the era when computer memory was measured in kilobytes, so that even storage of a large matrix was a problem. The only way to reduce the required memory was to reduce the number of degrees of freedom. Since discretization of the flow region (domain) requires fine grids, the decrease of number of degrees of freedom was sought in different versions of global weighted residual methods, called also spectral methods. These methods do not discretize the domain, and approximate solutions as truncated series of basis functions. It was shown by many authors that weighted residuals methods allow for a very significant reduction of degrees of freedom (see, e.g., Canuto *et al.,* 2006). Straight-forward application of these methods to incompressible fluid dynamics involves algebraic constraints that are related to the boundary conditions and the continuity equation. Removal of these constraints by an appropriate choice of basis functions would decrease the number of degrees of freedom even more. The construction of such basis functions, is the main topic of this text. For problems with periodic boundary conditions the choice of Fourier series is natural, and also allows for application of the fast Fourier transform when the non-linear terms are evaluated by the pseudo-spectral approach. However, the choice of basis functions is not so obvious for non-periodic boundary conditions. An obvious idea to linearly combine well-known functions, e.g., the Chebyshev polynomials, into expressions that satisfy linear and homogeneous boundary conditions of a problem appears in Orszag (1971a,b). To the best of the author's knowledge, this is the first appearance of this idea, at least in CFD. Many authors, including this one, rediscovered this way of constructing the basis functions (see Section 9).

The next step was to combine these linear superpositions into a two-dimensional divergence-free vector, for which Chebyshev polynomials of the first and second kinds fit very well (Gelfgat 1988, Gelfgat & Tanasawa, 1994). The Galerkin method based on divergence-free basis functions that satisfy all linear and homogeneous boundary conditions (LHBC in the further text), led to a noticeable reduction of the number of degrees of freedom, as was already shown in Gelfgat & Tanasawa (1994). Later, as computational power grew, so did the possibilities of applications of the method. A small number of degrees of freedom resulted in Jacobian matrices of small size, which could be treated numerically for the computation of steady flows, as well as for the solution of the eigenvalue problems associated with linear stability (Gelfgat et al., 1997, 1999a,b). Divergence free functions were extended to cylindrical coordinates, which allowed us to consider flow in the rotating disk – cylinder system and to obtain the first stability results (Gelfgat et al., 1996) for this configuration.

Since then a number of different problems addressing steady flows, multiplicity of states, and stability were solved for different flows in two-dimensional rectangular cavities and cylinders. The periodic circumferential coordinate in the cylindrical geometry allows one to study stability of an axisymmetric base flow with respect to three-dimensional perturbations. The linearized problem for the 3D disturbances separates for each circumferential Fourier mode, so that the final answer is obtained by consideration of several 2D-like problems. Subsequently, a considerable number of results were obtained for stability of flows in cylinders, which were driven by the rotation of boundaries, by buoyancy convection, and by magnetic field.

The first attempt to study 3D instability in Cartesian coordinates was carried out in Gelfgat (1999) for the Rayleigh-Bénard problem in a rectangular box, i.e., the stability of a quiescent fluid heated from below. No attempts were made by the author to study stability of fully developed 3D flow. At the same time, the three-dimensional bases found an unexpected application for visualization of incompressible 3D flows (Gelfgat, 2014).

In what follows we start from a brief description of the weighted residual and Galerkin method formulated for an incompressible fluid dynamics problem. Then the proposed way of constructing basis functions is explained, starting from a one-dimensional two-point problem. The treatment of inhomogeneities and boundary conditions singularities is explained. This follows by discussion of the resulting dynamical system and explanations of how it was treated in the cited works, as well as how it can be treated for fully 3D problems. After that, several illustrative examples are presented. Finally, a discussion of possible future studies is given.

## 2    The problem and the numerical method

We consider flow of an incompressible fluid in a two-dimensional rectangle $0 \leq x \leq A_x, 0 \leq y \leq A_y$, or in a three-dimensional box $0 \leq x \leq A_x, 0 \leq y \leq A_y, 0 \leq z \leq A_z$. The rectangular shape of the domain is the main restriction for all the following. Below we discuss how this restriction can be relaxed to a canonical shape, i.e., the domain bounded by the coordinate surfaces belonging to a system of orthogonal curvilinear coordinates. The momentum and continuity equations for velocity $\boldsymbol{v}$ and pressure $p$ read

$$\frac{\partial \boldsymbol{v}}{\partial t} + (\boldsymbol{v} \cdot \nabla)\boldsymbol{v} = -\nabla p + \frac{1}{Re}\Delta \boldsymbol{v} + \boldsymbol{f} , \qquad (2.1)$$

$$\nabla \cdot \boldsymbol{v} = 0 \quad , \qquad (2.2)$$

where $Re$ is the Reynolds number. We assume also that boundary conditions for all three velocity components are linear and homogeneous, e.g., the no-slip conditions on all boundaries. Equations (2.1)-(2.2) can be considered together with other scalar transport equations for, e.g., temperature and/or concentration, an example of which will be given below.

To formulate the Galerkin method we assume that the solution $\boldsymbol{v}$ belongs to a space $\boldsymbol{\mathcal{V}}$ of divergence-free vectors satisfying all the (linear and homogeneous) boundary conditions, LHBC. Assume that vectors $\{\boldsymbol{\varphi}_K\}_{K=1}^{\infty}$ form a basis in this space. Then the solution $\boldsymbol{v}$ can be represented as

$$\boldsymbol{v} = \sum_{K=1}^{\infty} c_K \boldsymbol{\varphi}_K . \qquad (2.3)$$

The coefficients $c_K$ can be obtained by evaluation of inner products of Eq. (2.3) with a basis vector $\boldsymbol{\varphi}_L$:

$$\langle \boldsymbol{v}, \boldsymbol{\varphi}_L \rangle = \sum_{K=1}^{\infty} c_K \langle \boldsymbol{\varphi}_K, \boldsymbol{\varphi}_L \rangle, \quad L = 1,2,3,\ldots \qquad (2.4)$$

Here we assumed that the space $\boldsymbol{\mathcal{V}}$ is supplied with an inner product $\langle \cdot,\cdot \rangle$, which is yet to be defined. If the functions $\{\boldsymbol{\varphi}_K\}_{k=1}^{\infty}$ form an orthogonal basis, then the coefficients $c_K$ are obtained as

$$c_K = \frac{\langle \boldsymbol{v}, \boldsymbol{\varphi}_K \rangle}{\langle \boldsymbol{\varphi}_K, \boldsymbol{\varphi}_K \rangle}, \quad K = 1,2,3,\ldots \qquad (2.5)$$

However, if the basis functions are not orthogonal, the expressions (2.4) form an infinite system of linear algebraic equations, which can be solved only up to a certain truncation. Keeping the first $N$ terms in the series (2.3) and defining a vector of coefficients $\boldsymbol{c} = \{c_1, c_2, c_3, \ldots, c_N\}$, the first $N$ coefficients can be obtained as

$$\boldsymbol{c} = G^{-1}\boldsymbol{f}, \quad G_{KL} = \langle \boldsymbol{\varphi}_K, \boldsymbol{\varphi}_L \rangle, \quad f_L = \langle \boldsymbol{v}, \boldsymbol{\varphi}_L \rangle. \qquad (2.6)$$

Here $G$ is the Gram matrix, and $N$ is the truncation number.

In the relations (2.4)-(2.6) we assumed that the vector $\boldsymbol{v}$ is known. If it is unknown, the coefficients $c_K$ can be obtained only approximately by minimization of the residual of the momentum equation. To show how they can be calculated, we first assume that the coefficients $c_K$ are time-dependent and the basis functions $\{\boldsymbol{\varphi}_K\}_{K=1}^{\infty}$ depend only on coordinate values. Then the representation (2.3) defines a time- and space-dependent function $\boldsymbol{v}$. Note that the continuity equation (2.2), as well as all the boundary conditions are already satisfied because $\boldsymbol{v}$ belongs to the space $\boldsymbol{\mathcal{V}}$. Clearly, the solution we are looking for also belongs to this space, so that the momentum equation is the only one to be solved. Now we rewrite the momentum equation (2.1) in two additional and equivalent forms

$$\frac{\partial \boldsymbol{v}}{\partial t} + (\boldsymbol{v} \cdot \nabla)\boldsymbol{v} + \nabla p - \frac{1}{Re}\Delta \boldsymbol{v} - \boldsymbol{f} = \boldsymbol{R} = \boldsymbol{0} \qquad (2.7)$$

$$\frac{\partial \boldsymbol{v}}{\partial t} = -\nabla p + \frac{1}{Re}\Delta \boldsymbol{v} - (\boldsymbol{v} \cdot \nabla)\boldsymbol{v} + \boldsymbol{f} \qquad (2.8)$$

Here $\boldsymbol{R}$ is the residual of the momentum equation. If $\boldsymbol{v}$ is the solution then $\boldsymbol{R} \equiv 0$. For a general case we assume that all possible residuals belong to a certain functional space $\boldsymbol{\mathcal{W}}$, which is also supplied with an inner product $\langle \cdot, \cdot \rangle_W$. Assume also that $\{\boldsymbol{\phi}_K\}_{K=1}^{\infty}$ is a basis in $\boldsymbol{\mathcal{W}}$. Then the requirement that the residual be zero is equivalent to the requirement that the residual $\boldsymbol{R}$ is orthogonal to all the basis functions $\boldsymbol{\phi}_K$, namely

$$\langle \boldsymbol{R}, \boldsymbol{\phi}_K \rangle_W = 0, \quad K = 1, 2, 3, \ldots \qquad (2.9)$$

The equations (2.9) form an infinite set of non-linear time-dependent ODEs that must be solved to find the time-dependent coefficients $c_K(t)$. However, this system of equations is not closed yet, since nothing is said about the pressure. To proceed, we assume that the pressure belongs to a space $\boldsymbol{\mathcal{S}}$ of scalar time-dependent functions, differentiable at least twice in the domain, with the basis $\{s_K\}_{K=1}^{\infty}$. The pressure is represented as

$$p = \sum_{K=1}^{\infty} d_K(t) s_K \ . \qquad (2.10)$$

The equation for pressure is formed in the standard way, by applying the divergence operator to both sides of the momentum equation (2.1), which yields

$$\Delta p = -div[(\boldsymbol{v} \cdot \nabla)\boldsymbol{v} - \boldsymbol{f}] \ . \qquad (2.11)$$

The boundary conditions required this equation will be discussed later. Note, that the velocity representation (2.3), even in the truncated form used below, guarantees zero velocity divergence.

Since the Laplacian and divergence operators are evaluated analytically, they commute in every approximate (truncated) formulation. We introduce the residual $D$ of the pressure equation

$$D = \Delta p + div[(\boldsymbol{v} \cdot \nabla)\boldsymbol{v} - \boldsymbol{f}]. \qquad (2.12)$$

For the residual $D$ we demand only piecewise continuity in all spatial directions, so that it formally belongs to another space of scalar functions. This space is denoted as $\mathcal{D}$, its basis as $\{q_K\}_{K=1}^{\infty}$, and the scalar product as $\langle \cdot, \cdot \rangle_D$. Now, for the solution of the problem represented by $\boldsymbol{v}$ and $p$, the scalar non-linear algebraic equations

$$\langle D, q_L \rangle_D = 0, \quad L = 1, 2, 3, \ldots \qquad (2.13)$$

must be satisfied. Equations (2.13) define the coefficients $d_K(t)$ and must be satisfied together with the equations (2.9). Note that these equations do not contain the time derivative and, therefore, are algebraic constraints for the ODEs (2.9).

Obviously, $\mathcal{V} \subseteq \mathcal{W}$ and $\mathcal{S} \subseteq \mathcal{D}$. To build a numerical procedure for obtaining the first $N_v$ and $N_p$ coefficients of the velocity and pressure series (2.3) and (2.10), we truncate the series together with the residual projection equations (2.9) and (2.13). This results in

$$\boldsymbol{v} \approx \sum_{K=1}^{N_v} c_K(t)\boldsymbol{\varphi}_K, \quad \langle \boldsymbol{R}, \boldsymbol{\phi}_L \rangle_W = 0, \quad L = 1, 2, 3, \ldots, N_v, \qquad (2.14)$$

$$p \approx \sum_{K=1}^{N_p} d_K(t) s_K, \quad \langle D, q_M \rangle_D = 0, \quad M = 1, 2, 3, \ldots, N_p, \qquad (2.15)$$

which defines $N_v$ non-linear ODEs and $N_p$ non-linear algebraic constraints for calculation of the time-dependent coefficients $c_K(t)$ and $d_K(t)$. This is called method of weighted residuals (Fletcher (1984), Boyd (2000), Canuto *et al.* (2006)). The ODEs and the algebraic constraints resulting from projections of the momentum and pressure equation residuals onto the basis functions are

$$\sum_{K=1}^{N_v} \dot{c}_K(t) \langle \boldsymbol{\varphi}_K, \boldsymbol{\phi}_M \rangle_W = -\sum_{K=1}^{N_v} \sum_{L=1}^{N_v} c_K(t) c_L(t) \langle (\boldsymbol{\varphi}_L \cdot \nabla) \boldsymbol{\varphi}_K, \boldsymbol{\phi}_M \rangle_W - \sum_{K=1}^{N_p} d_K(t) \langle \nabla s_K, \boldsymbol{\phi}_M \rangle_W +$$
$$\frac{1}{Re} \sum_{K=1}^{N_v} c_K(t) \langle \Delta \boldsymbol{\varphi}_K, \boldsymbol{\phi}_M \rangle_W + \langle \boldsymbol{f}, \boldsymbol{\phi}_M \rangle_W, \quad M = 1, 2, 3, \ldots, N_v \qquad (2.16)$$

$$\sum_{K=1}^{N_p} d_K(t) \langle \Delta s_K, q_J \rangle_D = -\sum_{K=1}^{N_v} \sum_{L=1}^{N_v} c_K(t) c_L(t) \langle \nabla \cdot (\boldsymbol{\varphi}_L \cdot \nabla) \boldsymbol{\varphi}_K, q_J \rangle_D + \langle \nabla \cdot \boldsymbol{f}, q_J \rangle_D \qquad (2.17)$$

$$J = 1, 2, 3, \ldots, N_p$$

Following common definitions, $\{\boldsymbol{\varphi}_K\}_{K=1}^{\infty}$ and $\{s_K\}_{K=1}^{\infty}$ are called coordinate basis systems, while $\{\boldsymbol{\phi}_K\}_{K=1}^{\infty}$ and $\{q_K\}_{K=1}^{\infty}$ - are called projection basis systems.

Note that the weighted residuals method can be formulated also for coordinate basis functions that do not satisfy some or all boundary conditions. Fletcher (1984) distinguishes between coordinate functions that satisfy only a (linear) differential equation, satisfy only boundary conditions, and do not satisfy anything, calling these three cases boundary, interior, and mixed, respectively. Within this classification, and noting that the equations are non-linear, we discuss mainly interior methods.

To arrive at the Galerkin formulation we assume that the boundary conditions do not explicitly involve time. Then the l.h.s. of Eq. (2.8) satisfies the LHBC of the velocity and is divergence-free. Then, the same can be said about the r.h.s. of Eq. (2.8), and finally about the residual of the momentum equation $R$ defined in Eq. (2.7). Thus, for time-independent boundary conditions, the residual $R$ belongs to the space $\mathcal{V}$, so that we can choose $\boldsymbol{\phi}_K = \boldsymbol{\varphi}_K$ for all $K$. Also assuming that the coordinate systems $\{\boldsymbol{\varphi}_K\}_{K=1}^{\infty}$ and $\{s_K\}_{K=1}^{\infty}$ are usually constructed from trigonometric functions or polynomials that are infinitely differentiable, the residual $D$ will also be differentiable infinite number of times. Since the physical problem does not specify any pressure boundary conditions, we can assume that both $p$ and $D$ belong to the same space $\mathcal{S}$, and we can choose $q_K = s_K$ for all $K$. Thus, we arrive at a particular version of the weighted residuals method, in which the coordinate and projection basis systems coincide. This version is known as the Galerkin or Boubnov – Galerkin method.

An obvious reason to choose the Galerkin method among all possible weighted residual formulations follows from the fact that $\mathcal{V} \subseteq \mathcal{W}$. By projecting onto a smaller space $\mathcal{V}$, we expect that with increase of the truncation number convergence will be faster. Another, less obvious and more profound reason follows from the definition of the inner products via volume integrals. For scalar and vector functions defined in the domain $V$ and on its boundary we define

$$\langle f(x,y,z), g(x,y,z) \rangle_\rho = \int_V \rho(x,y,z) f g \, dV, \qquad (2.18)$$

$$\langle \boldsymbol{u}(x,y,z), \boldsymbol{v}(x,y,z) \rangle_\rho = \int_V \rho(x,y,z) \boldsymbol{u} \cdot \boldsymbol{v} \, dV, \qquad (2.19)$$

where $\rho(x,y,z) > 0$ is the weight function. The simplest and the most robust formulation is obtained with the unit weight $\rho(x,y,z) = 1$, for which the inner products (2,18), (2.19) may also have some physical meaning. For example, the norm produced by (2.19) becomes twice the dimensionless kinetic energy. An important additional advantage of the unit weight follows from consideration of the inner product of the gradient of a scalar field $f(x,y,z)$ with a divergence free vector field $\boldsymbol{u}(x,y,z)$. Assume that $\Gamma$ is the domain boundary, $\boldsymbol{n}$ its normal, and that the component of $\boldsymbol{u}$ normal to the boundary vanishes. Keeping in mind that $\nabla \cdot \boldsymbol{u} = 0$ and $\boldsymbol{u} \cdot \boldsymbol{n} = 0$, we have

$$\langle \nabla f, \boldsymbol{u} \rangle_1 = \int_V \nabla f \cdot \boldsymbol{u} \, dV = \int_V [\nabla \cdot (f\boldsymbol{u}) - f \nabla \cdot \boldsymbol{u}] \, dV = \int_V \nabla \cdot (f\boldsymbol{u}) \, dV = \int_\Gamma f\boldsymbol{u} \cdot \boldsymbol{n} \, d\Gamma = 0 \quad (2.20)$$

Thus, if the velocity does not penetrate the boundary, and the inner product is chosen as in (2.19), the projection of the pressure gradient on all the velocity basis functions vanishes. This means that equation systems (2.16) and (2.17) separate, so that velocity can be calculated from (2.14) without any knowledge about pressure. Note that this calculation involves both minimization of the residual and summation of the truncated velocity series. The pressure then can be computed from (2.15) using the previously found velocity field. In this case Eqs. (2.16), which minimizes the residual by computing the time-dependent coefficients $c_K(t)$, become

$$\sum_{K=1}^{N_v} \dot{c}_K(t) \langle \boldsymbol{\varphi}_K, \boldsymbol{\varphi}_M \rangle_1 =$$
$$-\sum_{K=1}^{N_v} \sum_{L=1}^{N_v} c_K(t) c_L(t) \langle (\boldsymbol{\varphi}_L \cdot \nabla) \boldsymbol{\varphi}_K, \boldsymbol{\varphi}_M \rangle_1 + \frac{1}{Re} \sum_{K=1}^{N_v} c_K(t) \langle \Delta \boldsymbol{\varphi}_K, \boldsymbol{\varphi}_M \rangle_1 + \langle \boldsymbol{f}, \boldsymbol{\varphi}_M \rangle_1 \quad (2.21)$$

$$M = 1, 2, 3, \ldots, N_v$$

This is the formal Galerkin formulation for calculating an approximate solution of a problem. Note that the exclusion of pressure by the Galerkin projection (2.20) is not restricted to closed regions with non-penetrative boundaries. The inhomogeneity in the velocity boundary conditions can be removed by change of variables, which is discussed below in more detail.

To proceed we need to explain how to build basis functions, which are divergence-free and satisfy the whole set of LHBC. Then we will discuss how to handle inhomogeneous boundary conditions, curvilinear coordinates, and weight functions others than unity.

## 3  Basis functions

We start from the question of how to satisfy all the LHBC for a scalar unknown function. We assume that the basis functions for all three-dimensional time-dependent scalar variables, e.g., temperature and/or concentration, are represented as products of some one-dimensional bases. Thus for a function $\theta(x,y,z,t)$ defined in a box $0 \leq x \leq A_x, 0 \leq y \leq A_y, 0 \leq z \leq A_z$ we seek a representation in the form of the tensor (Kronecker) product of one-dimensional bases

$$\theta(x,y,z,t) = \sum_{k=1}^{N_z}\sum_{j=1}^{N_y}\sum_{i=1}^{N_x} d_{ijk}(t)f_i(x)g_j(y)h_k(z), \tag{3.1}$$

Here $d_{ijk}(t)$ are unknown time-dependent coefficients, $N_x, N_y, N_z$ are the truncation numbers specified in each spatial direction separately, and $f_i(x), g_j(y), h_k(z)$ are one-dimensional bases that must be defined in each direction. The three-dimensional basis corresponding to those defined in the previous chapter is

$$F_J(x,y,z) = f_i(x)g_j(y)h_k(z), \quad J = N_x[N_y(k-1)+j-1]+i \tag{3.2}$$

Starting from here we will use capital letters for the global indices, and small letters for the one-dimensional indices.

If all the boundary conditions for $\theta$ are linear and homogeneous, the functions $f_i(x), g_j(y)$ and $h_k(z)$ must satisfy the boundary conditions posed in the $x$-, $y$-, and $z$-directions, respectively. Assume that there are $M$ boundary conditions in the $x$-direction posed in the form

$$\sum_{l=0}^{L} \alpha_{ml} f^{(l)}(x_m) = 0, \quad m = 1, 2, \ldots, M. \tag{3.3}$$

Here $\alpha_{ml}$ are known coefficients, $l$ is the derivative number, and $x_m$ are the borders (that are parts of coordinate surfaces $x = x_m$) where the boundary conditions are posed. Note that $m$ is the number of a boundary condition and not a number of the point, since several boundary conditions can be defined at the same point, so that sometimes $x_{m_1} = x_{m_2}$ for $m_1 \neq m_2$. For the following we can also extend (3.3) by assuming that negative $l$ correspond to integrals, and that surfaces $x = x_m$ are not necessarily the boundary points, but also can lie inside the flow region, as it happens in a two-fluid example below. Now assume that a set of functions $\{s_k(x)\}_{k=1}^{\infty}$ forms a basis in, say, $C^{\infty}([0, A_x])$. This can be, for example, a trigonometric Fourier basis, or a set of orthogonal polynomials defined on $[0, A_x]$. For some reasons we choose this basis for representation of the solution, but the functions $s_k(x)$ do not satisfy the boundary conditions (3.3). We build an alternative basis by considering superpositions of $M+1$ consequent basis functions as

$$r_k(x) = \sum_{n=0}^{M} \beta_{nk} s_{k+n}(x) \tag{3.4}$$

Substituting $r_k(x)$ into the boundary conditions (3.3) we obtain

$$\sum_{n=0}^{M} \beta_{nk} r_k(x_m) = \sum_{l=0}^{L} \alpha_{ml} \sum_{n=0}^{M} \beta_{nk} s_{k+n}^{(l)}(x_m) = 0, \quad m = 1, 2, \ldots, M \tag{3.5}$$

For each $k$ the relations (3.5) form a system of $M$ equations that can be used to find $M+1$ coefficients $\beta_{nk}$. To make the equations solvable we choose $\beta_{0,k} = 1$, that leaves us with $M$ linear algebraic equations for $M$ remaining coefficients $\beta_{nk}$ for every fixed $k$. The matrix of this system will be regular if the boundary conditions (3.3) are independent. In this way, the functions $r_k(x)$ are fully defined and satisfy all the boundary conditions (3.3). Obviously they form a basis in the subspace $span\{r_k(x)\} \subset C^\infty([0, A_x])$, which contains functions that satisfy conditions (3.3) and are differentiable infinite number of times. In this way we form bases $f_i(x), g_j(y)$ and $h_k(z)$ for the representation (3.1).

In the following all the examples will be based on the Chebyshev polynomials of the first and the second type, $T_k(x)$ and $U_k(x)$, briefly described in the Appendix A. In the further text these polynomials will always be chosen as the basis $\{s_k(x)\}_{k=1}^\infty$.

### 3.1. *Example: Two-point boundary value problem*

Consider a two-point boundary value problem for $\theta(x)$ posed on the interval $0 \leq x \leq 1$ with the two ($M = 2$) boundary conditions

$$a_0\,\theta'(0) + b_0\,\theta(0) = 0, \quad a_1\,\theta'(1) + b_1\,\theta(1) = 0, \tag{3.1.1}$$

where $a_0$, $b_0$, $a_1$, $b_1$ are known coefficients. Recalling the Chebyshev polynomials and taking into account $\beta_{0,k} = 1$, our new basis functions are defined as

$$r_k(x) = \sum_{n=0}^{2} \beta_{nk} T_{k+n}(x) = T_k(x) + \beta_{1k} T_{k+1}(x) + \beta_{2k} T_{k+2}(x)\ . \tag{3.1.2}$$

The boundary values of the Chebyshev polynomials and their derivatives are given in Appendix A. Substituting $r_k(x)$ into the boundary conditions (3.1.1) and using Eqs. (A3) and (A5), we obtain two equations for the coefficients $\beta_{1k}$ and $\beta_{2k}$:

$$-\beta_{1k}[2a_0(k+1)^2 + b_0] + \beta_{2k}[2a_0(k+2)^2 + b_0] = -2a_0 k^2 - b_0 \tag{3.1.3}$$

$$\beta_{1k}[2a_1(k+1)^2 + b_1] + \beta_{2k}[2a_1(k+2)^2 + b_1] = -2a_1 k^2 - b_1 \tag{3.1.4}$$

These equation are easily solved analytically, which yields

$$\beta_{1k} = \frac{(2a_1 k^2 + b_1)[2a_0(k+2)^2 + b_0] - (2a_0 k^2 + b_0)[2a_1(k+2)^2 + b_1]}{[2a_0(k+1)^2 + b_0][2a_1(k+2)^2 + b_1] + [2a_1(k+1)^2 + b_1][2a_0(k+2)^2 + b_0]}, \tag{3.1.5}$$

$$\beta_{2k} = \frac{(2a_1 k^2 + b_1)[2a_0(k+1)^2 + b_0] + (2a_0 k^2 + b_0)[2a_1(k+1)^2 + b_1]}{[2a_0(k+1)^2 + b_0][2a_1(k+2)^2 + b_1] + [2a_1(k+1)^2 + b_1][2a_0(k+2)^2 + b_0]}, \tag{3.1.6}$$

and defines a new basis, whose satisfy both boundary conditions. This idea was introduced in Orszag (1971a) for two-point homogeneous Dirichlet boundary conditions, and in Orszag (1971b) for boundary conditions of the Orr-Sommerfeld equation. It was formalized for an arbitrary set of boundary conditions (3.1.1) in Gelfgat (1988). Note that for a similar problem defined on the interval $0 \leq x \leq A_x$ we need only to replace $x$ by $x/A_x$ in Eq. (3.1.2), which will not affect the expressions (3.1.5) and (3.1.6).

## 3.2. Two-dimensional divergence-free basis

To construct basis functions, which are two-dimensional divergence-free vectors satisfying all the boundary conditions, we start by constructing a divergence-free basis that does not yet involve any boundary conditions. Using the Chebyshev polynomials, and assuming the flow region is a rectangle $0 \leq x \leq A_x$, $0 \leq y \leq A_y$, we define

$$\boldsymbol{w}_{ij} = \begin{Bmatrix} w_{ij}^{(x)} \\ w_{ij}^{(y)} \end{Bmatrix} = \begin{Bmatrix} \frac{A_x}{2i} T_i\left(\frac{x}{A_x}\right) U_{j-1}\left(\frac{y}{A_y}\right) \\ -\frac{A_y}{2j} U_{i-1}\left(\frac{x}{A_x}\right) T_j\left(\frac{y}{A_y}\right) \end{Bmatrix}, \quad i,j = 1,2,3,\ldots \qquad (3.2.1)$$

$$\boldsymbol{w}_{0j} = \begin{Bmatrix} \frac{A_x}{2} U_{j-1}\left(\frac{y}{A_y}\right) \\ 0 \end{Bmatrix}, \quad \boldsymbol{w}_{i0} = \begin{Bmatrix} 0 \\ \frac{A_y}{2} U_{i-1}\left(\frac{x}{A_x}\right) \end{Bmatrix} \qquad (3.2.2)$$

Applying the relation (A2), it is easily seen that $\nabla \cdot \boldsymbol{w}_{ij} = 0$. Now, to implement the boundary conditions and to keep the divergence zero, we keep the *x*- and *y*- components of the vector dependent on each other, as in (3.2.1). To do so we implement all the velocity boundary conditions in the *x*-direction in the *x*-dependent part of the vector, and do the same in the *y*-direction. In the *x*-direction the boundary conditions are posed at $x = 0$ and $x = A_x$ for the two velocity components, so that we have four boundary conditions in total. The same is done in the *y*-direction. Thus, we construct the basis using linear superpositions of five (4+1) consecutive Chebyshev polynomials. This results in

$$\boldsymbol{\varphi}_{ij}(x,y) = \begin{Bmatrix} \frac{A_x}{2} \sum_{m=0}^{4} \frac{\sigma_{im}}{i+m} T_{i+m}\left(\frac{x}{A_x}\right) \sum_{l=0}^{4} \tau_{jl} U_{j+l-1}\left(\frac{y}{A_y}\right) \\ -\frac{A_y}{2} \sum_{m=0}^{4} \sigma_{im} U_{i+m-1}\left(\frac{x}{A_x}\right) \sum_{l=0}^{4} \frac{\tau_{jm}}{j+l} T_{j+l}\left(\frac{y}{A_y}\right) \end{Bmatrix} \qquad (3.2.3)$$

It is easy to check that $\nabla \cdot \boldsymbol{\varphi}_{ij} = 0$. The coefficients $\sigma_{im}$ and $\tau_{jm}$ must be found by substituting $\boldsymbol{\varphi}_{ij}$ into the boundary conditions. For example, assume that the rectangle has a stress free boundary at $y = A_y$, while all the other boundaries are no-slip. This leads to the following boundary conditions for the velocity $\boldsymbol{v}$:

$$\boldsymbol{v}(x=0,y) = \boldsymbol{v}(x=A_x,y) = \boldsymbol{v}(x,y=0) = 0, \quad v_y(x,y=A_y) = \frac{\partial v_x}{\partial y}(x,y=A_y) = 0 \quad (3.2.4)$$

The assignment $\sigma_{i0} = \tau_{i0} = 1$ and substitution of (3.2.3) into (3.2.4) yields

$$\sigma_{i1} = \sigma_{i3} = 0, \quad \sigma_{02} = -\frac{8}{3}, \quad \sigma_{04} = \frac{4}{3} \tag{3.2.5}$$

$$\sigma_{i2} = -\frac{i}{i+2} - \frac{(i+1)(i+4)^2}{i(i+2)(i+3)}, \quad \sigma_{i4} = \frac{(i+1)(i+4)}{i(i+3)}, \quad i > 0 \tag{3.2.6}$$

$$\tau_{01} = \frac{2}{7}, \quad \tau_{i1} = 2\frac{i^2+2i+1}{i^3+5i^2+7i}, \quad i > 0 \tag{3.2.7}$$

$$\tau_{02} = -\frac{16}{7}, \quad \tau_{i2} = -2\frac{i^4+8i^3+26i^2+40i+24}{i^4+8i^3+22i^2+21i}, \quad i > 0 \tag{3.2.8}$$

$$\tau_{03} = -\frac{6}{7}, \quad \tau_{i3} = -2\frac{i^2+4i+3}{i^3+5i^2+7i}, \quad i > 0 \tag{3.2.9}$$

$$\tau_{04} = \frac{4}{7}, \quad \tau_{i4} = \frac{i^4+8i^3+22i^2+27+12}{i^4+8i^3+22i^2+21i}, \quad i > 0 \tag{3.2.10}$$

### *3.3. Three-dimensional divergence-free basis*

Generalization of the 2D basis functions for a three-dimensional case is not straight-forward, since it is unclear how to produce divergence-free vectors, similar to those defined in (3.2.3), that will form a complete set of basis functions. To use a similar approach, we need to carry out several additional calculations.

Assume that $\boldsymbol{v} = (u, v, w)$ is a divergence-free three-dimensional vector that satisfies the no-slip conditions on all the boundaries of the rectangular box $0 \leq x \leq A_x, 0 \leq y \leq A_y, 0 \leq z \leq A_z$. Since $\nabla \cdot \boldsymbol{v} = \partial u/\partial x + \partial v/\partial y + \partial w/\partial z = 0$, so that $w = -\int (\partial u/\partial x + \partial v/\partial y)\, dz$, a 3D incompressible vector field can be decomposed as

$$\boldsymbol{v} = \begin{pmatrix} u \\ v \\ w \end{pmatrix} = \begin{pmatrix} u \\ 0 \\ w_1 \end{pmatrix} + \begin{pmatrix} 0 \\ v \\ w_2 \end{pmatrix}, \quad w_1 = -\int_0^z \frac{\partial u}{\partial x} dz, \quad w_2 = -\int_0^z \frac{\partial v}{\partial y} dz \tag{3.3.1}$$

Since both vectors $v^{(x,z)} = (u, 0, w_1)$ and $v^{(y,z)} = (0, v, w_2)$ are divergence-free, this decomposition shows that a divergence-free velocity field can be represented as superposition of two fields which have components only in the $(x,z)$ or $(y,z)$ planes. For each of the two vectors we can construct basis functions similar to (3.2.3)

$$\boldsymbol{\varphi}_{ijk}^{(x,z)}(x,y,z) = \begin{bmatrix} \frac{A_x}{2}\sum_{l=0}^{4}\frac{\hat{a}_{il}}{(i+l)}T_{i+l}\left(\frac{x}{A_x}\right)\sum_{m=0}^{4}\hat{b}_{jm}T_{j+m}\left(\frac{y}{A_y}\right)\sum_{n=0}^{\tilde{n}}\hat{c}_{kn}U_{k+n-1}\left(\frac{z}{A_z}\right) \\ 0 \\ -\frac{A_z}{2}\sum_{l=0}^{4}\hat{a}_{il}U_{i+l-1}\left(\frac{x}{A_x}\right)\sum_{m=0}^{4}\hat{b}_{jm}T_{j+m}\left(\frac{y}{A_y}\right)\sum_{n=0}^{\tilde{n}}\frac{\hat{c}_{kn}}{(k+n)}T_{k+n}\left(\frac{z}{A_z}\right) \end{bmatrix}$$
(3.3.2)

$$\boldsymbol{\varphi}_{ijk}^{(y,z)}(x,y,z) = \begin{bmatrix} 0 \\ \frac{A_y}{2}\sum_{l=0}^{4}\tilde{a}_{il}T_{i+l}\left(\frac{x}{A_x}\right)\sum_{m=0}^{4}\frac{\tilde{b}_{jm}}{(j+m)}T_{j+m}\left(\frac{y}{A_y}\right)\sum_{n=0}^{\tilde{n}}\tilde{c}_{kn}U_{k+n-1}\left(\frac{z}{A_z}\right) \\ -\frac{A_z}{2}\sum_{l=0}^{4}\tilde{a}_{il}T_{i+l}\left(\frac{x}{A_x}\right)\sum_{m=0}^{4}\tilde{b}_{jm}U_{j+m-1}\left(\frac{y}{A_y}\right)\sum_{n=0}^{\tilde{n}}\frac{\tilde{c}_{kn}}{(k+n)}T_{k+n}\left(\frac{z}{A_z}\right) \end{bmatrix}$$
(3.3.3)

As above, the coefficients $\hat{a}_{il}$, $\hat{b}_{jm}$, $\hat{c}_{kn}$, $\tilde{a}_{il}$, $\tilde{b}_{jm}$, and $\tilde{c}_{kn}$ are defined after substitution of the functions in the boundary conditions. Note that the number of polynomials included in the linear superpositions in $z$-direction, $\tilde{n} + 1$, is not yet defined. This is because at $z = A_z$ only the sum $w_1(x, y, A_z) + w_2(x, y, A_z)$ is zero, while the boundary values of $w_1(x, y, A_z)$ and $w_2(x, y, A_z)$ are not known. An example of such a decomposition can be found in Gelfgat (2014). Therefore there are only three boundary conditions in the $z$-direction to be satisfied by the basis functions $\boldsymbol{\varphi}_{ijk}^{(x,z)}$ and $\boldsymbol{\varphi}_{ijk}^{(y,z)}$, so that $\tilde{n} = 3$. It is still possible to use these bases, but to satisfy the boundary conditions for $w$ at $z = A_z$, additional algebraic constraints will be needed. Note that there is no such a problem if boundary conditions in the $z$-direction are periodic.

To remove the above algebraic constraint at $z = A_z$, we assume that $\tilde{n} = 4$ in (3.3.2) and (3.3.3), so that the functions $\boldsymbol{\varphi}_{ijk}^{(x,z)}$ and $\boldsymbol{\varphi}_{ijk}^{(y,z)}$ are divergence-free and satisfy all the boundary conditions. In this case, using (3.2.5) and (3.2.6),

$$\hat{a}_{il} = \tilde{a}_{il} = \hat{b}_{il} = \tilde{b}_{il} = \hat{c}_{il} = \tilde{c}_{il} = \sigma_{il}, \ \hat{b}_{j1} = \tilde{b}_{j1} = \hat{b}_{j3} = \tilde{b}_{j3} = \hat{b}_{j4} = \tilde{b}_{j4} = 0,$$

$$\hat{b}_{j2} = \tilde{b}_{j2} = -\frac{(j+2)^2}{j^2} \quad (3.3.4)$$

Projection of the solution $v$ onto $span\{\boldsymbol{\varphi}_{ijk}^{(x,z)}\}$ and $span\{\boldsymbol{\varphi}_{ijk}^{(y,z)}\}$ results in a vector $\tilde{v}$

$$\widetilde{\boldsymbol{v}} = \begin{pmatrix} \widetilde{u} \\ 0 \\ \widetilde{w}_1 \end{pmatrix} + \begin{pmatrix} 0 \\ \widetilde{v} \\ \widetilde{w}_2 \end{pmatrix}, \tag{3.3.5}$$

which is divergence-free, satisfies all the boundary conditions, but may not be a good approximation of $\boldsymbol{v}$ because the set of basis functions still is not complete. To complete the basis we notice that $span\{\boldsymbol{\varphi}_{ijk}^{(x,z)}\}$ and $span\{\boldsymbol{\varphi}_{ijk}^{(y,z)}\}$ project the velocity on the $(x,z)$ and $(y,z)$ planes, so that it is straight-forward to add projections on the $(x,y)$ planes as well. Thus, similarly to (3.3.2) and (3.3.3) we add another set of divergence-free basis functions satisfying all the boundary conditions

$$\boldsymbol{\varphi}_{ijk}^{(x,y)}(x,y,z) = \begin{bmatrix} \frac{A_x}{2} \sum_{l=0}^{4} \frac{\bar{a}_{il}}{(i+l)} T_{i+l}\left(\frac{x}{A_x}\right) \sum_{m=0}^{4} \bar{b}_{jm} U_{j+m-1}\left(\frac{y}{A_y}\right) \sum_{n=0}^{4} \bar{c}_{kn} T_{k+n}\left(\frac{z}{A_z}\right) \\ \frac{-A_y}{2} \sum_{l=0}^{4} \bar{a}_{il} U_{i+l-1}\left(\frac{x}{A_x}\right) \sum_{m=0}^{4} \frac{\bar{b}_{jm}}{(j+m)} T_{j+m}\left(\frac{y}{A_y}\right) \sum_{n=0}^{4} \bar{c}_{kn} T_{k+n}\left(\frac{z}{A_z}\right) \\ 0 \end{bmatrix} \tag{3.3.6}$$

Similarly to previous functions, for the no-slip boundary conditions posed on all boundaries,

$$\bar{a}_{il} = \bar{b}_{il} = \sigma_{il}, \quad \bar{c}_{k1} = \bar{c}_{k3} = \bar{c}_{k4} = 0, \quad \bar{c}_{k2} = -\frac{(k+2)^2}{k^2} \tag{3.3.7}$$

Finally, we seeek a three-dimensional solution of the form

$$\begin{aligned}
\boldsymbol{v} &\approx \sum_{i=0}^{L^{(x,y)}} \sum_{j=0}^{M^{(x,y)}} \sum_{k=0}^{N^{(x,y)}} C_{i,j,k}^{(x,y)}(t) \boldsymbol{\varphi}_{ijk}^{(x,y)}(x,y,z) + \\
&+ \sum_{i=0}^{L^{(x,z)}} \sum_{j=0}^{M^{(x,z)}} \sum_{k=0}^{N^{(x,z)}} C_{i,j,k}^{(x,z)}(t) \boldsymbol{\varphi}_{ijk}^{(x,z)}(x,y,z) + \\
&+ \sum_{i=0}^{L^{(y,z)}} \sum_{j=0}^{M^{(y,z)}} \sum_{k=0}^{N^{(y,z)}} C_{i,j,k}^{(y,z)}(t) \boldsymbol{\varphi}_{ijk}^{(y,z)}(x,y,z)
\end{aligned} \tag{3.3.8}$$

Projection of the residual of the momentum equation on all three sets of basis vectors yields three sets of ODEs for calculation of the three sets of time-dependent coefficients $C_{i,j,k}^{(x,y)}, C_{i,j,k}^{(x,z)}$, and $C_{i,j,k}^{(y,z)}$. Since the basis functions are not orthogonal it will be necessary to invert the Gram matrix that is formed as

$$G = \begin{bmatrix} \langle \boldsymbol{\varphi}_{ijk}^{(x,y)}, \boldsymbol{\varphi}_{pqr}^{(x,y)} \rangle & \langle \boldsymbol{\varphi}_{ijk}^{(x,y)}, \boldsymbol{\varphi}_{pqr}^{(x,z)} \rangle & \langle \boldsymbol{\varphi}_{ijk}^{(x,y)}, \boldsymbol{\varphi}_{pqr}^{(y,z)} \rangle \\ \langle \boldsymbol{\varphi}_{ijk}^{(x,z)}, \boldsymbol{\varphi}_{pqr}^{(x,y)} \rangle & \langle \boldsymbol{\varphi}_{ijk}^{(x,z)}, \boldsymbol{\varphi}_{pqr}^{(x,z)} \rangle & \langle \boldsymbol{\varphi}_{ijk}^{(x,z)}, \boldsymbol{\varphi}_{pqr}^{(y,z)} \rangle \\ \langle \boldsymbol{\varphi}_{ijk}^{(y,x)}, \boldsymbol{\varphi}_{pqr}^{(x,y)} \rangle & \langle \boldsymbol{\varphi}_{ijk}^{(y,x)}, \boldsymbol{\varphi}_{pqr}^{(x,z)} \rangle & \langle \boldsymbol{\varphi}_{ijk}^{(y,z)}, \boldsymbol{\varphi}_{pqr}^{(y,z)} \rangle \end{bmatrix} \qquad (3.3.9)$$

A simple numerical test for the no-slip boundary conditions and equal truncation numbers (starting from 4 and larger) in each direction and for each set of the functions shows that the Gram matrix is singular. This means that some of the functions are linearly dependent and must be excluded. Based on the above discussion, we see that in the case of periodic boundary conditions in the *z*-direction, all the set (3.3.6) can be omitted. However, some functions of this set must be added in the case of no-slip boundaries. This shows that the complete set of linearly independent basis functions differs for different boundary conditions. Unfortunately, the author could not arrive at a rigorous mathematical procedure that would determine which functions must be excluded at certain boundary conditions. At the same time, a simple numerical experiment can be helpful.

Considering no-slip conditions at all the boundaries, we varied $N^{(y,z)}$ in the last sum of Eq. (3.3.7). In other words, we varied the truncation number in the *z*-direction for the functions defined in (3.3.6) only. We found that by taking $N^{(y,z)} = 0$, i.e., one basis function in the *z*-direction, we obtain a regular Gram matrix. Increase of $N^{(y,z)}$ to $N^{(y,z)} \geq 1$, makes the Gram function singular. Furthermore, taking a single basis function in the *z*-direction, with the third index $k \geq 1$, which means polynomials of larger degrees in (3.3.6), also results in a singular Gram matrix. This shows that the addition of the first polynomial in the *z*-direction, corresponding to $N^{(y,z)} = 0$, is essential, while the others can be omitted. Using (3.3.7), this first polynomial is $8(z - z^2)$. Returning to the sets (3.3.2) and (3.3.3) with the coefficients defined in (3.3.4), we observe that the polynomials corresponding to the *x*- and *y*- vector components start form the second degree, while those corresponding to the z-component start from the third one. Thus, the missing second-order polynomial must be added with the help of another set (3.3.6).

### *3.4. Basis functions in cylindrical and other curvilinear coordinates*

Consider flow in a cylinder with radius $R$ and height $H$. The whole problem is defined now in the cylindrical coordinates $(r, \theta, z)$, $0 \leq r \leq R$, $0 \leq \theta \leq 2\pi$, $0 \leq z \leq H$. Using $2\pi$-periodicity in the azimuthal direction we represent the flow as a Fourier series

$$\boldsymbol{v} = \sum_{k=-\infty}^{\infty} \boldsymbol{v}_k(r,z) exp(ik\theta) , \qquad (3.4.1)$$

so that the basis functions in the $\theta$-direction are complex exponents $exp(ik\theta)$. The continuity equation for $\boldsymbol{v}_k(r,z) = \big(u_k(r,z), v_k(r,z), w_k(r,z)\big)$ is

$$\frac{1}{r}\frac{\partial(ru_k)}{\partial r} + \frac{ik}{r}v_k + \frac{\partial w_k}{\partial z} = 0 \qquad (3.4.2)$$

Here we must distinguish between the axisymmetric case and axisymmetric Fourier mode, for which $k = 0$, and all the other $k \neq 0$ modes. Axisymmetric flow (or the axisymmetric mode) is represented by a single set of the basis functions, which is built similarly to the above 2D Cartesian case, but taking into account the continuity equation (3.4.2). Note that if an axisymmetric flow has also a non-zero azimuthal component, the latter can be treated as a scalar function. Then the axisymmetric vector basis should include only the radial and axial components. An example of such basis, successfully used in several studies (see below) is

$$\boldsymbol{U}_{ij}(r,z) = \begin{Bmatrix} u_0 \\ w_0 \end{Bmatrix} = \begin{Bmatrix} \frac{1}{2}\frac{r}{R}\sum_{m=0}^{4} \frac{\sigma_{im}}{i+m} T_{i+m}\left(\frac{r}{R}\right) \sum_{m=0}^{4} \mu_{jm} U_{j+m-1}\left(\frac{z}{A_z}\right) \\ -\frac{A_z}{2}\sum_{m=0}^{4} \sigma_{im}\widetilde{U}_{i+m-1}\left(\frac{r}{R}\right) \sum_{m=0}^{4} \frac{\mu_{jm}}{j+m} T_{j+m}\left(\frac{z}{A_z}\right) \end{Bmatrix}, \qquad (3.4.3)$$

where $\widetilde{U}_n\left(\frac{r}{R}\right) = T_{n+1}\left(\frac{r}{R}\right) + r(n+1)U_n\left(\frac{r}{R}\right)$. As above, the zero divergence of $\boldsymbol{U}_{ij}(r,z)$ follows from Eq. (A2), and the coefficients $\sigma_{im}$ and $\mu_{jm}$ are obtained by substitution of $\boldsymbol{U}_{ij}(r,z)$ in the boundary conditions. Note also that the $r$-component of velocity vanishes at the polar axis $r = 0$, so that for flow in a full cylinder only three conditions in the radial direction must be additionally satisfied. If the domain is a cylindrical layer (e.g., Taylor-Couette flow) then the polar axis is cut out and one remains with the four boundary conditions, as in the Cartesian case.

Now consider Fourier modes of (3.4.1) corresponding to $k \neq 0$. Using the same idea as in Eq. (3.1.1) we observe that

$$\boldsymbol{v}_k = \begin{pmatrix} u_k \\ v_k \\ w_k \end{pmatrix} = \begin{pmatrix} u_k \\ -\frac{1}{ik}\frac{\partial(ru_k)}{\partial r} \\ 0 \end{pmatrix} + \begin{pmatrix} 0 \\ -\frac{r}{ik}\frac{\partial w_k}{\partial z} \\ w_k \end{pmatrix}. \qquad (3.4.4)$$

Obviously, the sum of two azimuthal components of this relation satisfies the boundary conditions for $\boldsymbol{v}_k$. It is not clear, however, whether each of them satisfies the boundary conditions separately. The author is not sure that this can be proved for a general case, but it can be easily examined for no-slip conditions at $r = R$ and $z = 0, H$. Since $w_k(r = R) = 0$ the derivative $\partial w_k/\partial z$ also vanishes at $r = R$. Since the sum of two azimuthal components vanishes at $r = R$, the azimuthal component of the first vector of the r.h.s also vanishes there, so that each component satisfies boundary conditions in the radial direction. Similarly, we consider $\partial(ru_k)/\partial r$ at $z = 0, H$ and conclude that it vanishes there because $u_k(r, 0) = u_k(r, H) = 0$. Then also the second azimuthal component vanishes at $z = 0, H$, and all the no-slip boundary conditions for the azimuthal velocity, as well as the axis condition, are satisfied by each r.h.s. vector of (3.4.4) separately. Thus, considering flows with no-slip cylindrical boundaries, we can decompose $v_{k \neq 0}$ into two independent bases, so that the whole flow will be represented as

$$v = \sum_{i=0}^{M_r} \sum_{j=0}^{M_z} A_{ij}(t) U_{ij}(r,z) + \sum_{\substack{k=-\infty \\ k \neq 0}}^{k=+\infty} \left\{ \sum_{i=0}^{N_r} \sum_{j=0}^{N_z} [B_{ij}^k(t) V_{ij}(r,z) + C_{ij}^k(t) W_{ij}(r,z)] \right\} \exp(ik\theta) \quad (3.4.5)$$

Here the functions $U_{ij}(r, z)$ represent the axisymmetric part of the flow and are defined in (3.4.3). The functions $V_{ij}(r, z)$ and $W_{ij}(r, z)$ represent the two vectors in r.h.s. of (3.4.4) and are defined as

$$\boldsymbol{V}_{ij}(r,z) = \begin{Bmatrix} -ik\left(\frac{r}{R}\right)^q \sum_{m=0}^{4} \bar{\sigma}_{im} T_{i+m}\left(\frac{r}{R}\right) \sum_{l=0}^{4} \bar{\mu}_{jl} T_{j+l}\left(\frac{z}{A_z}\right) \\ \\ \sum_{m=0}^{4} \bar{\sigma}_{im} \bar{U}_{i+m-1}\left(\frac{r}{R}\right) \sum_{l=0}^{4} \bar{\mu}_{jl} T_{j+l}\left(\frac{z}{A_z}\right) \\ \\ \end{Bmatrix}, \quad (3.4.5)$$

$$\boldsymbol{W}_{ij}(r,z) = \begin{Bmatrix} \left(\frac{r}{R}\right)^2 \sum_{m=0}^{4} \bar{\bar{\sigma}}_{im} T_{i+m}\left(\frac{r}{R}\right) \sum_{l=0}^{4} \bar{\bar{\mu}}_{jl} U_{j+l-1}\left(\frac{z}{A_z}\right) \\ \\ -\frac{ikA_z}{2} r \sum_{m=0}^{4} \bar{\bar{\sigma}}_{im} T_{i+m}\left(\frac{r}{R}\right) \sum_{l=0}^{4} \frac{\bar{\bar{\mu}}_{jl}}{j+l} T_{j+ll}\left(\frac{z}{A_z}\right) \\ \\ \end{Bmatrix}, \quad (3.4.6)$$

Here $q = 0$ for $|k| = 1$ and $q = 1$ for $|k| > 1$, $\bar{U}_n(x) = (q+1)r^q T_n(x) + 2nr^{q+1} U_{n-1}(x)$. Again, the zero divergence of the functions $\boldsymbol{V}_{ij}(r, z) exp(ik\theta)$ and $\boldsymbol{W}_{ij}(r, z) exp(ik\theta)$ follow from Eqs. (3.4.2) and (A2), and the coefficients $\bar{\sigma}_{im}$, $\bar{\mu}_{jm}$, $\bar{\bar{\sigma}}_{im}$, and $\bar{\bar{\mu}}_{jm}$ are defined after substitution of (3.4.5) and (3.4.6) in the boundary conditions. The integer parameter $q$ appears because of different boundary conditions posed for $|k| = 1$ and $|k| \neq 1$ at the polar axis (Canuto, et al., 2006; Gelfgat et al. 1999), which are

At $r = 0$: $u_{k=0} = 0, v_{k=0} = 0, \frac{\partial w_{k=0}}{\partial r} = 0$

$$u_{k=\pm 1} \neq 0, v_{k=\pm 1} \neq 0, w_{k=\pm 1} = 0 \qquad (3.4.7)$$

$$u_{|k|>1} = 0, v_{|k|>1} = 0, w_{|k|>1} = 0$$

For no-slip conditions at the top, bottom and sidewall of the cylinder the coefficients $\sigma_{im}$, $\mu_{jm}$, $\bar{\sigma}_{im}, \bar{\mu}_{jm}, \bar{\bar{\sigma}}_{im}$, and $\bar{\bar{\mu}}_{jm}$ are defined as

$$\sigma_{i1} = -\frac{i^3+7i^2+15i+9}{i^3+6i^2+12i+8}, \quad \sigma_{i2} = -\frac{i^2}{(i+2)^2}, \quad \sigma_{i3} = \frac{i^3+3i^2+3i+1}{i^3+6i^2+12i+8}, \quad \sigma_{i4} = 0 \qquad (3.4.8)$$

$$\bar{\sigma}_{i1} = -\frac{4(i+1)}{2i+3}, \quad \bar{\sigma}_{i2} = \frac{2i+1}{2i+3}, \quad \bar{\sigma}_{i3} = \frac{2i+1}{4(i+2)}, \quad \bar{\sigma}_{i4} = 0 \qquad (3.4.9)$$

$$\bar{\bar{\sigma}}_{i1} = -1, \quad \bar{\bar{\sigma}}_{i2} = 0, \quad \bar{\bar{\sigma}}_{i3} = 0, \quad \bar{\bar{\sigma}}_{i4} = 0 \qquad (3.4.10)$$

$$\mu_{i1} = \mu_{i3} = \bar{\bar{\mu}}_{i1} = \bar{\bar{\mu}}_{i3} = 0, \quad \mu_{02} = \bar{\bar{\mu}}_{02} = -\frac{8}{3}, \quad \mu_{04} = \bar{\bar{\mu}}_{04} = \frac{4}{3}; \qquad (3.4.11)$$

$$\mu_{i2} = \bar{\bar{\mu}}_{i2} = -\frac{i}{i+2} - \frac{(i+1)(i+4)^2}{i(i+2)(i+3)}, \quad \mu_{i4} = \bar{\bar{\mu}}_{i4} = \frac{(i+1)(i+4)}{i(i+3)}, \quad i > 0 \qquad (3.4.12)$$

$$\bar{\mu}_{i1} = \bar{\mu}_{i3} = \bar{\mu}_{i4} = 0, \quad \bar{\mu}_{i2} = -1 \qquad (3.4.13)$$

This example of divergence-free basis functions built for cylindrical geometries also shows how construction of a divergence free basis satisfying all the boundary conditions can be approached in other orthogonal coordinate systems. The process can be noticeably simplified if two periodic spatial coordinates are involved in the formulation of the problem. In these cases only a one-dimensional basis will be needed. Alternatively, the divergence-free basis in cylindrical and spherical coordinates with two periodic directions can be defined as in Dumas & Leonard (1994) or Meseguer & Melibovsky (2007).

## 4. Inhomogeneous boundary conditions

If the problem has inhomogeneous boundary conditions, they can be included as algebraic constraints. A better method is a change of variables so that all inhomogeneities are moved from the boundary conditions to the equations. Then the boundary conditions become homogeneous, and the corresponding basis functions can be built as in Section 3.4. We assume that all the boundary conditions are linear. Such change of variables can use analytic, as well as numerically calculated functions. Several examples of this are briefly described below.

The simplest example for the change of variables is convection in a box, which has constant temperatures at the opposite sides, while all the other boundaries are perfectly thermally conducting or perfectly insulated. In the first case the temperature varies linearly along these boundaries, while in the second case temperature derivative normal o the boundary must vanish. These boundary conditions are satisfied by the linear temperature profile, which corresponds to the temperature distribution in a purely conducting case. For example, if for the dimensionless temperature $\theta(x,y,z)$, the boundary conditions in the $x$-direction are $\theta(x = 0, y, z) = 1$, $\theta(x = 1, y, z) = 0$, the change of variable is $\theta = (1 - x) + \tilde{\theta}(x, y, z)$. The function $(1 - x)$ satisfies all homogeneous and inhomogeneous boundary conditions, so that all the boundary conditions for new unknown function $\tilde{\theta}(x, y, z)$ are homogeneous. This change of variables was applied in all cited works of Gelfgat that treated convection in rectangular cavities starting from Gelfgat & Tanasawa (1994).

The inhomogeneities can be excluded from the boundary conditions by extracting a known analytical function from the solution only if the boundary conditions are continuous, including continuity at the corners of the computational domain. Another example of this is parabolic heating of a vertical cylinder that was considered in Gelfgat et al. (2000, 2001b). The dimensionless temperature at the cylindrical sidewall was prescribed as $\theta(r = 1, z) = f(z) = z(1 - z/A)$, $0 \leq z \leq A$, and was zero at the top and bottom, $z = 0, A$. Since the function $f(z)$ vanishes at the top and the bottom, the simplest change of variables in this case is $\theta = f(z) + \tilde{\theta}(r, z)$, which was applied in the mentioned studies.

Clearly, when the boundary conditions are discontinuous, use of a simple analytic function for the change of variables becomes problematic. Such a function can be built, for example, as a solution of Laplace equation with discontinuous boundary conditions. This analytic solution will suffer from Gibbs phenomenon which may destroy the convergence of the whole process. On the other hand, a low-order numerical solution can smooth the discontinuity up to an acceptable level, as in many calculations of lid-driven cavity flow, however this will take us too far from our Chebyshev-polynomial-based Galerkin approach. Thus, for calculation of the lid-driven cavity flow in Gelfgat (2005) we used analytically smoothed boundary condition, then solved the Stokes problem with the smoothed boundary conditions, and then used the Stokes problem solution for a change of variables. The Stokes problem was solved using the same Galerkin approach.

In the studies of swirling flows in a cylinder with rotating lid, as well as independently rotating top, bottom and sidewall of the cylinder (Gelfgat et al., 1996a,b, 2001; Marques et al., 2003) we solved the Stokes problem for the azimuthal velocity component. A similar Galerkin method in scalar formulation was applied. The boundary conditions with discontinuities in the corners were included as additional algebraic constraints by adding collocation points along the boundaries.

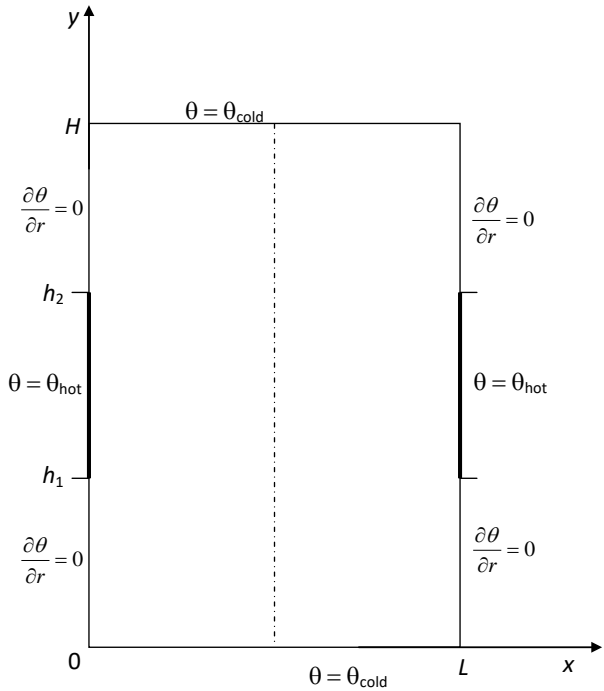

Fig. 1. Sketch of the temperature boundary conditions of the problem of Erenburg et al. (2003).

A more complicated case was treated in Erenburg et al. (2003). There we considered convection in a rectangular cavity with partially heated and partially insulated sidewall as sketched in Fig. 1. All the boundaries are no-slip. The dimensionless boundary conditions for the temperature are (here $A = H/L$, $a_1 = h_1/L$, $a_2 = h_2/L$)

$$\theta(x; y = 0, A) = 0, \quad (4.1)$$

$$\theta(x = 0,1; a_1 \leq y \leq a_2) = 1, \quad (4.2)$$

$$\frac{\partial \theta}{\partial x}(x = 0,1; y < a_1 \text{ or } y > a_2) = 0. \quad (4.3)$$

To arrive at a formulation with continuous and homogeneous boundary conditions on all the boundaries for a single unknown function, we decompose the temperature into two functions

$$\theta(x, y, t) = \Theta(x, y, t) + \tilde{\theta}(x, y, t) \quad (4.4)$$

where $\tilde{\theta}(x, y, t)$ is the new unknown function for which a continuous set of boundary conditions is required, i.e.,

$$\tilde{\theta}(x; y = 0, A) = 0, \quad \frac{\partial \tilde{\theta}}{\partial x}(x = 0,1; y) = 0 \quad (4.5)$$

The function $\Theta(x, y, t)$ is used to adjust the boundary conditions for $\theta(x, y, t)$ to (4.1)-4.3). Therefore, the boundary conditions for $\Theta(x, y, t)$ are

$$\Theta(x; y = 0, A) = 0, \tag{4.6}$$

$$\Theta(x = 0,1; a_1 \leq y \leq a_2) = 1 - \tilde{\theta}, \tag{4.7}$$

$$\frac{\partial \Theta}{\partial x}(x = 0,1; y < a_1 \text{ or } y > a_2) = 0 \tag{4.8}$$

To avoid the appearance of an additional source term in the energy equation we also require that $\Theta(x, y, t)$ be a solution of the Laplace equation,

$$\Delta \Theta = 0. \tag{4.9}$$

The solution of problem (4.6)-(4.9) can be represented as

$$\Theta(x, y, t) = \Theta_0(x, y, t) + \Theta_1(x, y, t) \tag{4.10}$$

where $\Theta_0(x, y, t)$ is the part of the solution of (4.6)-(4.9) corresponding to $\tilde{\theta} = 0$ and $\Theta_1(x, y, t)$ is the part dependent on $\tilde{\theta}$. Both functions $\Theta_0$ and $\Theta_1$ are calculated numerically by the global Galerkin method inside the cavity and collocation points at the sidewalls. Obviously, the part $\Theta_0(x, y, t)$ is defined by the geometry of the problem only, and is time-independent, so that it must be calculated only once. The problem formulation for $\Theta_1(x, y, t)$ is straight-forward, and its solution can be presented as $\Theta_1 = L^{-1}\tilde{\theta}$, where $L$ is the operator defining the problem, and approximated by its Galerkin/collocation projection. The representation of the temperature (4.4) now becomes

$$\theta(x, y, t) = \Theta(x, y, t) + \tilde{\theta}(x, y, t) = \Theta_0 + (L^{-1} + I)\tilde{\theta}, \tag{4.11}$$

where $I$ is the identity operator. The energy equation becomes

$$(L^{-1} + I)\frac{\partial \tilde{\theta}}{\partial t} + (\boldsymbol{v} \cdot \nabla)(L^{-1} + I)\tilde{\theta} = \frac{1}{Pr}\Delta(L^{-1} + I)\tilde{\theta} - (\boldsymbol{v} \cdot \nabla)\Theta_0. \tag{4.12}$$

Thus, after the function $\Theta_0$ and the operator $(L^{-1} + I)$ are calculated, the remaining problem for $\tilde{\theta}$ is defined with the homogeneous boundary conditions (4.5) only. Other details can be found in Erenburg et al. (2003).

## 5. Basis functions for two-fluid flow and boundary conditions at liquid-liquid interface

Here we give an example of basis functions that were used to calculate a two-phase flow with a capillary interface separating two liquids. A two-fluid Dean flow sketched in Fig. 2 was considered in Gelfgat et al. (2001d). The two fluids occupy adjacent thin cylindrical layers, $a \leq r \leq a + b$ and $a + b \leq r \leq a + d$, respectively, with the assumption $\bar{a} = a/d \gg 1$. The two fluids are separated by the border $r = a + b$. The base flow in both fluids is driven by a constant azimuthal pressure gradient. Note that while the pressure is then a multi-valued function of $\theta$, $\partial p/\partial \theta$ is not and can be considered as an imposed bulk force This formulation is an extension of the classical Dean (1928) problem to two-fluid system and is a simplified model of flow in a curved channel. Here we leave all the details concerning the evaluation of the base flow and the formulation of the stability problem to Gelfgat et al. (2001d), and focus only on the boundary conditions and incorporation of them into the basis functions. The three-dimensional velocity perturbation in cylindrical coordinates is defined as $\boldsymbol{v} = \big(u(x), v(x), w(x)\big)exp(\lambda t + in\theta + ikz)$. For convenience, we define a new dimensionless coordinate $x = (r-a)/d$, and define $\bar{b} = (b-a)/d$. The dimensionless amplitude of the interface perturbation is $\bar{\delta}$. Indices 1 and 2 denote the variables in each sublayer, $\rho_{12} = \rho_1/\rho_2$ and $\mu_{12} = \mu_1/\mu_2$ is the ratio of densities and viscosities, respectively. After the axial velocity $w$ and the pressure $p$ are eliminated, the conditions for the radial and azimuthal components at the bounding surfaces and the interface read

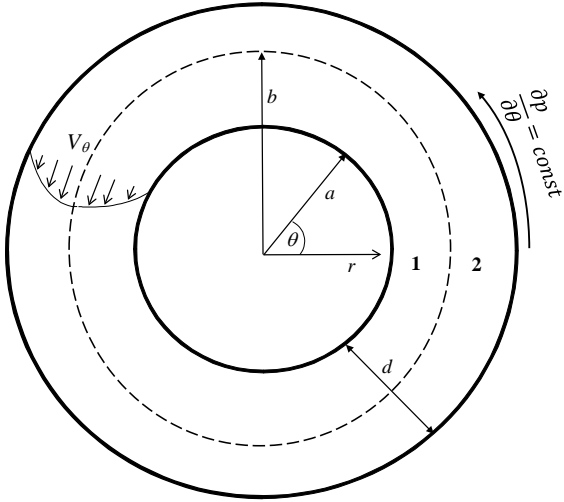

Fig. 2. Sketch of the two-fluid Dean flow problem

$$x = 0: \quad u_1 = v_1 = \frac{du_1}{dx} = 0 \quad (5.1)$$

$$x = 1: \quad u_2 = v_2 = \frac{du_2}{dx} = 0 \quad (5.2)$$

$$x = \bar{b}: \quad u_1 = u_2, \ v_1 = v_2 \quad (5.3)$$

$$\frac{du_1}{dx} = \frac{du_2}{dx} \quad (5.4)$$

$$\frac{dv_1}{dx} = \mu_{12}\frac{dv_2}{dx} \quad (5.5)$$

$$\frac{d^2u_1}{dx^2} + k^2 u_1 = \mu_{12}\left[\frac{d^2u_2}{dx^2} + k^2 u_2\right] \quad (5.6)$$

$$\lambda\bar{\delta} = u_1 - \frac{in}{\bar{b}}V\bar{\delta} \quad (5.7)$$

$$\lambda\left[\rho_{12}\frac{du_2}{dx} - \frac{du_1}{dx}\right] = \frac{\bar{\delta}}{\bar{a}}k^2(\rho_{12}-1)V^2 +$$

$$+ \frac{1}{Re}\left[\frac{d^2}{dx^2} - 2\frac{d}{dx} - k^2\right](\mu_{12}u_2 - u_1) - \rho_{12}\frac{in}{\bar{a}}V\frac{d}{dx}(\rho_{12}u_2 - u_1) - \frac{k^2}{We}\left[\frac{1-n^2}{\bar{b}^2} - k^2\right]\bar{\delta} \quad (5.8)$$

To construct basis functions, we start from non-deformable interface. In this case we add $u_1 = u_2 = 0$ to the boundary condition (5.3) and omit (5.7) and (5.8). Then the unknowns $u(x)$ and $v(x)$ are approximated by truncated series

$$u(x) = \sum_{m=1}^{N} c_m(t)\,\psi_m(x), \qquad v(x) = \sum_{l=1}^{N} d_m(t)\,\varphi_m(x), \tag{5.9}$$

where

$$\varphi_m(x) = \begin{cases} \sum_{l=0}^{2} \alpha_{ml}^{(1)} T_{m+l}\left(\frac{x}{\bar{b}}\right), & 0 \leq x \leq \bar{b} \\ \\ \sum_{l=0}^{2} \alpha_{ml}^{(2)} T_{m+l}\left(\frac{x-\bar{b}}{1-\bar{b}}\right), & \bar{b} \leq x \leq 1 \end{cases}, \tag{5.10}$$

$$\psi_m(x) = \begin{cases} \sum_{l=0}^{4} \beta_{ml}^{(1)} T_{m+l}\left(\frac{x}{\bar{b}}\right), & 0 \leq x \leq \bar{b} \\ \\ \sum_{l=0}^{4} \beta_{ml}^{(2)} T_{m+l}\left(\frac{x-\bar{b}}{1-\bar{b}}\right), & \bar{b} \leq x \leq 1 \end{cases}. \tag{5.11}$$

Here the superscripts (1) and (2) denote the sublayers. The coefficients $\alpha_{ml}^{(1)}, \alpha_{ml}^{(2)}, \beta_{ml}^{(1)}$, and $\beta_{ml}^{(2)}$ are obtained after substitution of the basis functions (5.11) into the boundary conditions (5.1)-(5.6). The inner product is defined as

$$\langle f, g \rangle = \int_0^1 f(x)\,g(x)dx = \int_0^{\bar{b}} f(x)\,g(x)dx + \int_{\bar{b}}^1 f(x)\,g(x)dx, \tag{5.12}$$

and after application of the Galerkin method, the time-dependent coefficients $c_m(t)$ and $d_m(t)$ are the same for the whole domain.

For the linear stability problem with deformable interface the solution is represented as

$$u(x) = c_0(t)\phi(x) + \sum_{m=1}^{N} c_m(t)\,\psi_m(x), \qquad v(x) = \sum_{m=1}^{N} d_m(t)\,\varphi_m(x). \tag{5.13}$$

The bases $\varphi_m(x)$ and $\psi_m(x)$ remain unchanged. An additional function $\phi(x)$ is introduced to satisfy the boundary conditions for a deformable interface. Its choice is arbitrary. In our case we define it as

$$\phi(x) = \begin{cases} \sum_{l=0}^{4} \gamma_{ql}^{(1)} T_{q+l}\left(\frac{x}{\bar{b}}\right), & 0 \leq x \leq \bar{b} \\ \\ \sum_{l=0}^{4} \gamma_{ql}^{(2)} T_{q+l}\left(\frac{x-\bar{b}}{1-\bar{b}}\right), & \bar{b} \leq x \leq 1 \end{cases}. \tag{5.14}$$

The coefficients $\gamma_{ql}^{(1)}$ and $\gamma_{ql}^{(2)}$ are used to satisfy the boundary conditions (5.1), (5.2), (5.4) and (5.6) subject to the normalization condition $\phi(\bar{b}) = 1$. Choice of the index $q$ is arbitrary, e.g., $q = 0$.

With the normalization condition $\phi(\bar{b}) = 1$ applied, the coefficient $c_0$ can be interpreted as the amplitude of the radial velocity at the deformed interface. This coefficient, and the interface amplitude $\bar{\delta}$, are defined by the two remaining boundary conditions (5.7) and (5.8). Thus, the Galerkin projections of the governing equations together with the boundary conditions (5.7) and (5.8) form a closed algebraic system for calculation of the coefficients $c_m$ and $d_m$ together with two additional unknown scalars $d_m$ and $\bar{\delta}$. The stability problem reduces to a generalized eigenvalue problem. In Gelfgat et al. (2001d) coefficients of the basis functions $\alpha_{ml}^{(1)}, \alpha_{ml}^{(2)}, \beta_{ml}^{(1)}, \beta_{ml}^{(2)}, \gamma_{ql}^{(1)}$, and $\gamma_{ql}^{(2)}$ were obtained by means of computer algebra.

## 6. The dynamical ODEs system for time-dependent coefficients

### *6.1. General expressions to be used coding the calculations*

In the following we assume that all the inner products are defined with unit weight. We also assume that all the necessary changes of variables are made, so that the boundary conditions of all the unknown vector and scalar fields are linear and homogeneous. Then, after the basis functions are constructed, and the Galerkin projections are made, and the pressure is eliminated by Eq. (2.20), we arrive at an ODE system (2.21) for the time-dependent coefficients. We store all the unknown time-dependent coefficients of the problem in the vector $\boldsymbol{X}(t) = \{X_I(t)\}_{I=1}^{N}$, where $N$ is the total number of scalar unknowns (degrees of freedom). Note that the vector $\boldsymbol{X}(t)$ contains velocity coefficients, as well as coefficients of all the other possible unknowns, e.g., temperature and/or concentration, excluding pressure. Then the resulting dynamic ODE system has the following form (the Einstein summation rule is assumed)

$$G_{IJ}\dot{X}_J = L_{IJ}X_J + N_{IJK}X_JX_K + F_I \quad . \tag{6.1.1}$$

Here $G_{IJ}$ is the Gram matrix, $L_{IJ}$, $N_{IJK}$, and $F_I$ are projections of the linear, nonlinear and bulk force of the momentum equation (2.7), respectively. The transport equation for temperature, concentration, electric and magnetic fields, etc., arrive to similar ODE systems that can be complicated by non-linear terms not belonging to the material derivative. This dynamical system has several nice properties that follow from the fact that the basis functions satisfy all boundary conditions, and are divergence-free. From Green's theorems for a scalar function and for a divergence-free velocity

$$\langle \Delta\theta, \theta \rangle = -\langle \nabla\theta, \nabla\theta \rangle, \quad \langle \Delta\boldsymbol{v}, \boldsymbol{v} \rangle = -\langle \nabla \times \boldsymbol{v}, \nabla \times \boldsymbol{v} \rangle. \tag{6.1.2}$$

It follows that the matrices $L_{IJ}$ corresponding to the dissipative terms are symmetric and negative definite independently on the truncation number. Furthermore, from the conservation properties

$$\langle (\boldsymbol{v} \cdot \nabla)\boldsymbol{v}, \boldsymbol{v} \rangle = 0, \qquad \langle (\boldsymbol{v} \cdot \nabla)\theta, \theta \rangle = 0 \tag{6.1.3}$$

it follows that

$$N_{IJK} X_I X_J X_K = 0 \tag{6.1.4}$$

for any truncation number. This means that the non-linear term always conserves the momentum. Additionally, these properties yield a very convenient tool for code debugging.

Computation of steady state flows reduces to an algebraic system of quadratic equations

$$L_{IJ} X_J + N_{IJK} X_J X_K + F_I = 0 \ , \tag{6.1.5}$$

for which we do not need to consider the Gram matrix. The application of Newton iteration is straightforward and requires computation and inversion of the Jacobian matrix

$$\mathfrak{J}_{IJ} = L_{IJ} + (N_{IJK} + N_{IKJ}) X_K \ . \tag{6.1.6}$$

Linear stability analysis of the calculated steady states reduces to computation of the eigenvalues of another Jacobian matrix, which includes the inverted Gram matrix

$$\widehat{\mathfrak{J}} Y = \lambda Y, \quad \widehat{\mathfrak{J}}_{IJ} = G_{IM}^{-1} [L_{MJ} + (N_{MJK} + N_{MKJ}) X_K] = \widehat{L}_{IJ} + (\widehat{N}_{IJK} + \widehat{N}_{IKJ}) X_K \ . \tag{6.1.7}$$

The inverse of the Gram matrix is also needed for straightforward time integration, for which the dynamical system (6.1) has the form

$$\dot{X}_I = G_{IM}^{-1} [L_{MJ} X_J + N_{MJK} X_J X_K + F_M] = \widehat{L}_{IJ} X_J + \widehat{N}_{IJK} X_J X_K + \widehat{F}_I \ , \tag{6.1.8}$$

where matrices multiplied by $G^{-1}$ are denoted by $\widehat{\phantom{x}}$.

Explicit representation of the dynamical system (6.1.8) allows one to perform additional analytical evaluations required by weakly non-linear analysis of bifurcations. This was implemented for the Hopf bifurcation in Gelfgat et al. (1996) and Gelfgat (2004). Assume that with increasing Reynolds number, at $Re = Re_{cr}$, a complex conjugate pair $\Lambda = \pm i\omega_0$ of leading eigenvalues of the problem (6.1.7) crosses the imaginary axis. Then the instability sets in as a Hopf bifurcation if all the conditions of the Hopf theorem hold, which is the most common case. We are interested in an asymptotic approximation of the oscillation period and amplitude at small super-criticality. Assume that $\boldsymbol{X}^0$ is the steady state at the critical point, and $\boldsymbol{U}$ and $\boldsymbol{V}$ are the left and right eigenvectors corresponding to the eigenvalue $\Lambda = i\omega_0$. We also denote the r.h.s. of the dynamic system (6.8) as $\boldsymbol{F}(\boldsymbol{X}; Re) = \dot{\boldsymbol{X}}$. Then, according to Hassard et al. (1981), the oscillating state, i.e., the limit cycle, is approximated as

$$Re = Re_{cr} + \mu_1\varepsilon^2 + O(\varepsilon^4) \qquad (6.1.9)$$

$$\tau(Re) = \frac{2\pi}{\omega_0}[1 + \tau_1\varepsilon^2 + O(\varepsilon^4)] \qquad (6.1.10)$$

$$\boldsymbol{X}(t, Re) = X^0(Re_{cr}) + \varepsilon Real\left[\boldsymbol{V}exp\left(\frac{2\pi i}{\tau}\right)\right] + O(\varepsilon^2) \qquad (6.1.11)$$

Here $\varepsilon$ is a formal positive parameter, $Re - Re_{cr}$ is the super-criticality, $\tau$ is the oscillation period, and $\boldsymbol{X}(t, Re)$ is the asymptotic oscillatory solution of the ODE system (6.1.8) for the Reynolds number defined in (6.1.9). This asymptotic expansion is defined by two parameters $\mu_1$ and $\tau_1$, which are calculated using the following procedu (Hassard et al., 1981)

$$\mu_1 = -\frac{Real(\sigma)}{\alpha_r}, \quad \tau_1 = -\frac{1}{\omega_0}[Im(\sigma) + \mu_1\alpha_i], \quad \alpha = \alpha_r + i\alpha_i = \left(\frac{d\Lambda}{dRe}\right)_{Re=Re_{cr}}, \qquad (6.1.12)$$

The parameter $\sigma$ is obtained as

$$\sigma = \frac{1}{2}H_{21} + \frac{1}{2\omega_0}\left[g_{20}g_{21} - 2|g_{11}|^2 - \frac{1}{3}|g_{02}|^2\right] \qquad (6.1.13)$$

$$g_{20} = 2\boldsymbol{U}^T\boldsymbol{f}_{20}, \qquad g_{02} = 2\boldsymbol{U}^T\bar{\boldsymbol{f}}_{20}, \qquad g_{11} = 2\boldsymbol{U}^T\boldsymbol{f}_{11}. \qquad (6.1.14)$$

The vectors $\boldsymbol{f}_{11}$ and $\boldsymbol{f}_{20}$ are the second derivatives of the r.h.s., and $H_{21}$ is the third derivative in the complex plane $C, \xi \in C$:

$$\boldsymbol{f}_{20} = \left\{\frac{\partial^2}{\partial\xi^2}\boldsymbol{F}\xi^2[X^0 + Real(\boldsymbol{V}\xi); Re_{cr}]\right\}_{\xi=0}, \quad \boldsymbol{f}_{11} = \left\{\frac{\partial^2}{\partial\xi\partial\bar\xi}\boldsymbol{F}[X^0 + Real(\boldsymbol{V}\xi); Re_{cr}]\right\}_{\xi=0}$$

$$(6.1.15)$$

$$H_{21} = \left\{\frac{\partial^3}{\partial\partial\bar\xi}\left(2\boldsymbol{U}^T\boldsymbol{F}[X^0 + Real(\boldsymbol{V}\xi + \boldsymbol{\varrho}_{20}\xi^2 + \boldsymbol{\varrho}_{11}\xi\bar\xi); Re_{cr}]\right)\right\}_{\xi=0}, \qquad (6.1.16)$$

and the vectors $\boldsymbol{\varrho}_{20}$ and $\boldsymbol{\varrho}_{11}$ are solutions of the following systems of linear algebraic equations ($\underline{I}$ is the identity matrix and $\otimes$ denotes the Kronecker product)

$$\widehat{\mathfrak{J}}\boldsymbol{\varrho}_{11} = -\boldsymbol{h}_{11}, \quad [\widehat{\mathfrak{J}} - 2i\omega_0\underline{I}]\boldsymbol{\varrho}_{11} = -\boldsymbol{h}_{20}, \quad \boldsymbol{h}_{ij} = [\underline{I} - Real(\boldsymbol{V}\otimes\boldsymbol{U}^T)]\boldsymbol{f}_{ij} \qquad (6.1.17)$$

The eigenvalue derivative (6.1.12) can be easily evaluated numerically. However, the most difficult part of this calculation is evaluation of the second and the third derivatives of the right hand side of the ODE system (6.1.15) and (6.1.16). These derivatives must be evaluated in the complex plane. The number of degrees of freedom is large, so that accurate enough differentiation by finite differences that will involve small increments of $\xi$ can be problematic. However, the explicit form of (6.1.8) allows for analytical calculation of the derivatives. The result is

$$f_{20,k} = \frac{1}{2}\left[N_{KIJ}\left(V_I^{(r)}V_J^{(r)} - V^{(i)}V_J^{(i)}\right) + iN_{KIJ}\left(V_I^{(r)}V_J^{(i)} - V_I^{(i)}V_J^{(r)}\right)\right], \qquad (6.1.18)$$

$$f_{11,k} = \frac{1}{2} N_{kij} \left( V_I^{(r)} V_J^{(r)} + V^{(i)} V_J^{(i)} \right) \quad , \tag{6.1.19}$$

$$G_{21} = \frac{1}{2} U_i \left( N_{ijm} + N_{imj} \right) \left( 2\varrho_{11,j} V_m + 2\varrho_{20,j} \bar{V}_m \right) \quad , \tag{6.1.20}$$

where superscripts $(r)$ and $(i)$ denote real and imaginary parts, respectively. These analytical expressions allow one to compute the asymptotic expansions (6.1.9)-(6.1.11) without significant loss of accuracy. The CPU-time requirements for such calculations are comparable with the calculation of two steady state solutions and their spectra. The sign of $\mu_1$ determines whether the bifurcation is sub- or super-critical.

## *6.2. Main computational difficulties*

All the computations described in the previous section, are restricted by two main difficulties. To describe these, we note that in all the studies where this method was successfully applied, the number of basis functions in one direction was between 20 and 70. Therefore, to make some estimates, we will address three types of truncation in one direction with 30, 60 and 100 basis functions for both two- and three-dimensional problems.

The first difficulty is connected with the Gram matrix $G$. In two-dimensional and quasi two-dimensional cases, e.g., 3D instability of an axisymmetric base flow, there is no problem storing the Gram matrix in memory, nor in computing its inverse. In fact, even with the truncation number 100, the Gram matrix consists of blocks of order $100^2=10^4$, which leads to an order of $10^8$ non-zero entries, which fits in several Gb memory. This matrix, which is symmetric and positive definite, can be inverted by Choleski decomposition, which is much faster than Gaussian elimination. This inverse must be computed only once for each specific problem and can be stored on disk. However, for all the calculations, except the Newton iterations, the r.h.s. of dynamic system must be multiplied by $G^{-1}$. If the truncation number is small, the matrices $\hat{L} = G^{-1}L$ and $\hat{N} = G^{-1}N$ in the dynamical system (6.1.7) can be computed and stored before other heavy computations begin. At larger truncation number the storage of matrix $\hat{N}$ becomes impossible (see below), which makes the time-dependent calculations too slow. At the same time, the stability analysis, as well as the weakly non-linear analysis, require only a few, usually less than 10, evaluations of the Jacobian matrix $\widehat{\mathfrak{I}}$, thus making the computational process affordable.

Treatment of the Gram matrix in a three-dimensional formulation is much more difficult. Here the largest block to be inverted has order of $((2M)^3)^2$ elements, where $M$ is the truncation number. Storage of such large matrices becomes problematic already at truncation numbers $M \geq 30$, and is unaffordable at $M = 100$. Thus, among all the possibilities described, only steady state calculations are possible. A solution for this can be orthonormalization of the set of the basis functions, which is discussed below.

The second difficulty is the numerical evaluation of non-linear term in (6.1.5) and (6.1.7), which requires the order of $((2M)^3)^3$ multiplications, assuming that the matrix $\widehat{N}$ is stored and evaluation of its terms does not require additional operations. Again, it can be affordable for the steady state, stability, and weakly non-linear calculations when 2D and quasi-2D problems are solved, because all these require very few evaluations of the r.h.s. In the fully 3D cases it seems to be not feasible, unless some additional evaluations of the non-linear terms are performed.

## 6.3. Treatment of non-linear terms

As a simplest example of handling the non-linear terms and avoiding too many multiplications, we consider the Burgers equation in the interval $0 \leq x \leq 1$

$$\frac{\partial u}{\partial t} + u \frac{\partial u}{\partial x} = \nu \frac{\partial^2 u}{\partial x^2}, \quad u(0,t) = u(1,t) = 0 \ . \tag{6.3.1}$$

The initial condition used in Gelfgat (2006) was $u(x,0) = sin(2\pi x) + sin(\pi x)/2$. We seek the solution as a truncated series

$$u(x,t) = \sum_{i=0}^{M-1} c_i(t)\varphi_i(x), \quad \varphi_i(x) = T_i(x) - T_{i+2}(x), \tag{6.3.2}$$

where the basis functions $\varphi_i(x)$ are constructed as linear superpositions of the Chebyshev polynomials to satisfy the boundary conditions (see Eqs. (A3)). After the Galerkin projections are applied we obtain the ODE system (6.1.1), where coefficients $c_i(t)$ are stored in the vector $\boldsymbol{X}$. For this problem the matrices in (6.1.1) are defined as

$$G_{ij} = \langle \varphi_j, \varphi_i \rangle, \ L_{ij} = \langle \varphi_j'', \varphi_i \rangle, \ F_i = 0, \ N_{ijk} = \langle \varphi_j \varphi_k', \varphi_i \rangle, \tag{6.3.3}$$

and evaluation of the non-linear term requires $M^3$ operations.

To reduce the number of multiplications, we notice that $\varphi_j \varphi_k'$ is a polynomial of order $j + k + 1$ that satisfies the boundary conditions of the problem, so that we can express it as a series of $\varphi_i(x)$:

$$\varphi_j(x)\varphi_k'(x) = \sum_{l=0}^{j+k-1} b_{ljk}\, \varphi_l(x) \tag{6.3.4}$$

The coefficients $b_{ljk}$ can be evaluated analytically using Eqs. (A7) and (A12) in the following way

$$\varphi_j(x)\varphi'_k(x) = [T_j(x) - T_{j+2}(x)][T'_k(x) - T'_{k+2}(x)] = \quad (6.3.5)$$

$$= [T_j(x) - T_{j+2}(x)]\left[4k \sum_{p=0}^{[(k-1)/2]} a_{k-1-2p} T_{k-1-2p}(x) - 4(k+2) \sum_{p=0}^{[(k+1)/2]} a_{k-1-2p} T_{k-1-2p}(x)\right]$$

$$= 2k \sum_{p=0}^{[(k-1)/2]} a_{k-1-2p}\{[T_{k-1-2p-j}(x) + T_{k-1-2p+j}(x)] - [T_{k-1-2p-j-2}(x) + T_{k-1-2p+j+2}(x)]\}$$

$$-2(k+2) \sum_{p=0}^{[(k+1)/2]} a_{k-1-2p}\{[T_{k-1-2p-j}(x)+T_{k-1-2p+j}(x)] - [T_{k-1-2p-j-2}(x)+T_{k-1-2p+j+2}(x)]\}$$

$$= 2k \sum_{p=0}^{[(k-1)/2]} a_{k-1-2p}\{T_{k-1-2p-j}(x) - T_{k-1-2p-j-2}(x) + T_{k-1-2p+j}(x) - T_{k-1-2p+j+2}(x)\}$$

$$-2(k+2) \sum_{p=0}^{[(k+1)/2]} a_{k-1-2p}\{T_{k-1-2p-j}(x) - T_{k-1-2p-j-2}(x) + T_{k-1-2p+j}(x) - T_{k-1-2p+j+2}(x)\}$$

$$= 2k \sum_{p=0}^{[(k-1)/2]} a_{k-1-2p}\{-\varphi_{k-1-2p-j-2}(x) + \varphi_{k-1-2p+j}(x)\}$$

$$-2(k+2) \sum_{p=0}^{[(k+1)/2]} a_{k-1-2p}\{-\varphi_{k-1-2p-j-2}(x) + \varphi_{k-1-2p+j}(x)\}$$

$$= -4 \sum_{p=0}^{[(k-1)/2]} a_{k-1-2p}\{-\varphi_{k-1-2p-j-2}(x) + \varphi_{k-1-2p+j}(x)\}$$

$$-2(k+2) \sum_{p=[(k-1)/2]+1}^{[(k+1)/2]} a_{k-1-2p}\{-\varphi_{k-1-2p-j-2}(x) + \varphi_{k-1-2p+j}(x)\}$$

Defining additionally $\varphi_{k<0} = 0$, noticing that $[(k-1)/2] + 1 = [(k+1)/2]$, and comparing the above result with (6.3.4) we observe that the coefficients $b_{ljk}$ can be assembled by the following procedure. Starting from $b_{ljk} = 0$,

$$b_{k-1-2[(k+1)/2]-j-2} = b_{k-1-2[(k+1)/2]-j-2} + 2(k+2)a_{k-1-2[(k+1)/2]}, \quad (6.3.6)$$
$$if \ k - 1 - 2[(k+1)/2] - j - 2 \geq 0$$

$$b_{k-1-2[(k+1)/2]+j} = b_{k-1-2[(k+1)/2]+j} - 2(k+2)a_{k-1-2[(k+1)/2]},$$
$$\text{if } k-1-2[(k+1)/2]+j \geq 0$$

For $p = 0$ to $p = [(k-1)/2]$:

$$b_{k-1-2p-j-2} = b_{k-1-2p-j-2} + 4a_{k-1-2p}, \quad \text{if } k-1-2p-j-2 \geq 0$$
$$b_{k-1-2p+j} = b_{k-1-2p+j} - 4a_{k-1-2p}, \quad \text{if } k-1-2p+j \geq 0$$

Now, using (6.3.4), we form a new set of time-dependent coefficients:

$$\sum_{j,k=0}^{M-1} c_j(t) c_k(t)\, \varphi_j(x) \varphi_k'(x) = \sum_{m=0}^{2(M-1)} C_m(t)\, \varphi_m(x), \tag{6.3.7}$$

$$C_m(t) = c_j(t) c_k(t) \sum_{l=0}^{j+k-1} (b_{ljk} + b_{lkj}), \quad m = j+k \tag{6.3.8}$$

And finally,

$$\sum_{j,k=0}^{M-1} N_{ijk} c_j(t) c_k(t) = \sum_{j,k=0}^{M-1} \langle \varphi_j \varphi_k', \varphi_i \rangle c_j(t) c_k(t) = \sum_{m=0}^{2(M-1)} C_m(t) \langle \varphi_m, \varphi_i \rangle \tag{6.3.9}$$

Now, we can estimate the number of multiplications required. The coefficients $b_{ljk}$, and the sums in (6.3.8) depend only on the basis functions and, therefore, can be computed only once in the beginning of the computational process. Computation of the coefficients $C_m(t)$ requires $2M^2$ multiplications and is the most CPU-time consuming part. Then, evaluation of (6.3.9) requires $2(M-1)$ multiplications providing that all the inner products are pre-computed. Since the operations in (6.3.8) and (6.3.9) are easily scalable, vectorization and/or parallel computing can additionally speed up the calculations.

Returning to the non-linear terms of momentum equation, we observe that in the case of no-slip conditions the non-linear terms $u\, \partial u/\partial x, v\, \partial u/\partial y, w \partial u/\partial z$, etc., satisfy the no-slip boundary conditions for $u, v$, and $w$, respectively. Thus, these terms can also be decomposed into series of appropriate basis functions, which will lead to a similar decrease in the number of required multiplications. When boundary conditions are more complicated, e.g., a stress-free boundary, one can add additional functions in which only boundary conditions satisfied by the non-linear terms are implemented. Alternatively, regardless of boundary conditions, the non-linear terms can be represented as Chebyshev polynomial series, which will also decrease the number of multiplications.

### 6.4. Orthogonalization and other polynomial bases

In this section we address two questions: does (i) orthogonalization of the basis or (ii) another polynomial basis change the final result? The answer is "no", but it requires some additional explanations.

After choosing the truncation number we seek the solution in the form of (3.3.8). The solution belongs to the linear space $\mathcal{L} = span\left\{\boldsymbol{\varphi}_{ijk}^{(x,y)}, \boldsymbol{\varphi}_{ijk}^{(x,z)}, \boldsymbol{\varphi}_{ijk}^{(y,z)}\right\}$ as in (3.3.8). This space consists of divergence-free vectors that satisfy all the LHBC of the problem, and their components are polynomials of maximum order $max[L^{(x,y)}, L^{(x,y)}, L^{(y,z)}] + 4$ in the $x$-direction, with similar expressions in the two other directions. We denote the order of space $\mathcal{L}$ as $N_\mathcal{L}$ and store all the basis functions of (3.3.8) in a set of vectors $\boldsymbol{Q} = \{\boldsymbol{q}_i\}_{i=1}^{N_\mathcal{L}}$. Clearly, this set forms a basis in $\mathcal{L}$, $\mathcal{L} = span\{\boldsymbol{Q}\}$. Assume now another basis in $\mathcal{L}$, denoted as $\widehat{\boldsymbol{Q}} = \{\widehat{\boldsymbol{q}}_i\}_{i=1}^{N_\mathcal{L}}$. The connection between the two bases is given by a matrix $\mathcal{B}$ of order $N_\mathcal{L}$ as

$$\widehat{\boldsymbol{Q}} = \mathcal{B}\boldsymbol{Q}, \quad \boldsymbol{Q} = \mathcal{B}^{-1}\widehat{\boldsymbol{Q}}. \tag{6.4.1}$$

Elements of the matrix $\mathcal{B}$ are solutions of the following system of linear algebraic equations (summation over repeating indices is assumed)

$$\langle \widehat{\boldsymbol{q}}_i, \widehat{\boldsymbol{q}}_j \rangle = \mathcal{B}_{ik} \langle \boldsymbol{q}_k, \widehat{\boldsymbol{q}}_j \rangle \tag{6.4.2}$$

Equation (6.4.1) can be interpreted as a transformation to another polynomial basis, a particular case of which is an orthonormal polynomial basis. In this case the matrix $\mathcal{B}$ is then the operator of the Gram-Schmidt or Householder orthogonalization procedure. Assume now that projection of the solution $\boldsymbol{v}$ on each of the bases is described by coefficients $X_i$ and $\widehat{X}_i$. Since these coefficients describe the orthogonal projection of the vector $\boldsymbol{v}$ onto the same space, the result of projection onto either basis must be identical, i.e.,

$$v \approx X_i \boldsymbol{q}_i = \widehat{X}_i \widehat{\boldsymbol{q}}_i = \widehat{X}_i \mathcal{B}_{ik} \boldsymbol{q}_k, \tag{6.4.3}$$

from which it follows that

$$X_i = \widehat{X}_i \mathcal{B}_{ik}, \quad \boldsymbol{X} = \mathcal{B}^T \widehat{\boldsymbol{X}}, \quad \widehat{\boldsymbol{X}} = (\mathcal{B}^T)^{-1} \boldsymbol{X} = (\mathcal{B}^{-1})^T \boldsymbol{X} \tag{6.4.4}$$

The Galerkin procedure applied with either of the two bases will result in two different ODE systems similar to (6.1.1):

$$G_{ij}\dot{X}_j = L_{ij}X_j + N_{ijk}X_jX_k + F_i, \quad \widehat{G}_{ij}\dot{\widehat{X}}_j = \widehat{L}_{ij}\widehat{X}_j + \widehat{N}_{ijk}\widehat{X}_j\widehat{X}_k + \widehat{F}_i, \tag{6.4.5}$$

where

$$G_{ij} = \langle \boldsymbol{q}_j, \boldsymbol{q}_i \rangle, \ L_{ij} = \frac{1}{Re}\langle \Delta\boldsymbol{q}_j, \boldsymbol{q}_i \rangle, \ N_{ijk} = \langle (\boldsymbol{q}_j \cdot \nabla)\boldsymbol{q}_k, \boldsymbol{q}_i \rangle, \ F_i = \langle \boldsymbol{f}, \boldsymbol{q}_i \rangle \tag{6.4.6}$$

$$\hat{G}_{ij} = \langle \hat{\boldsymbol{q}}_j, \hat{\boldsymbol{q}}_i \rangle, \quad \hat{L}_{ij} = \frac{1}{Re}\langle \Delta\hat{\boldsymbol{q}}_j, \hat{\boldsymbol{q}}_i \rangle, \quad \hat{N}_{ijk} = \langle (\hat{\boldsymbol{q}}_j \cdot \nabla)\hat{\boldsymbol{q}}_k, \hat{\boldsymbol{q}}_i \rangle, \quad \hat{F}_i = \langle \boldsymbol{f}, \hat{\boldsymbol{q}}_i \rangle \qquad (6.4.7)$$

Note that the two ODE systems in (6.4.5) describe the orthogonal projection of the residual on the same linear space $\mathcal{L}$, so that the result again must be identical. However, it is not clear yet whether the coefficients $X_i$ and $\hat{X}_i$ yielded by solution of the two systems will be connected via Eq. (6.4.4). Let us evaluate how the matrices in (6.4.6) and (6.4.7) are connected.

$$\hat{F}_i = \langle \boldsymbol{f}, \hat{\boldsymbol{q}}_i \rangle = \langle \boldsymbol{f}, \mathcal{B}_{ip}\boldsymbol{q}_p \rangle = \mathcal{B}_{ip}F_p \quad, \qquad (6.4.8)$$

$$\hat{G}_{ij} = \langle \hat{\boldsymbol{q}}_j, \hat{\boldsymbol{q}}_i \rangle = \langle \mathcal{B}_{jk}\boldsymbol{q}_k, \mathcal{B}_{ip}\boldsymbol{q}_p \rangle = \mathcal{B}_{jk}\mathcal{B}_{ip}G_{kp} \quad, \qquad (6.4.9)$$

$$\hat{L}_{ij} = \frac{1}{Re}\langle \Delta\hat{\boldsymbol{q}}_j, \hat{\boldsymbol{q}}_i \rangle = \frac{1}{Re}\langle \mathcal{B}_{jk}\Delta\boldsymbol{q}_k, \mathcal{B}_{ip}\boldsymbol{q}_p \rangle = \mathcal{B}_{jk}\mathcal{B}_{ip}L_{kp} \quad, \qquad (6.4.10)$$

$$\hat{N}_{ijk} = \langle (\hat{\boldsymbol{q}}_j \cdot \nabla)\hat{\boldsymbol{q}}_k, \hat{\boldsymbol{q}}_i \rangle = \langle (\mathcal{B}_{jm}\boldsymbol{q}_m \cdot \nabla)\mathcal{B}_{kl}\boldsymbol{q}_l, \mathcal{B}_{ip}\boldsymbol{q}_p \rangle = \mathcal{B}_{jm}\mathcal{B}_{kl}\mathcal{B}_{ip}N_{pml} \quad. \qquad (6.4.11)$$

Consider now how all the terms of the second system of (6.4.5) are expressed via matrices and unknowns of the first system

$$\hat{G}_{ij}\dot{\hat{X}}_j = \mathcal{B}_{jk}\mathcal{B}_{ip}G_{kp}\mathcal{B}_{jq}^{-1}\dot{X}_q = \mathcal{B}_{ip}G_{kp}\dot{X}_k \quad, \qquad (6.4.12)$$

$$\hat{L}_{ij}\hat{X}_j = \mathcal{B}_{jk}\mathcal{B}_{ip}L_{kp}\mathcal{B}_{jq}^{-1}X_q = \mathcal{B}_{ip}L_{kp}X_k \quad, \qquad (6.4.13)$$

$$\hat{N}_{ijk}\hat{X}_j\hat{X}_k = \mathcal{B}_{jm}\mathcal{B}_{kl}\mathcal{B}_{ip}N_{pml}\mathcal{B}_{jq}^{-1}X_q\mathcal{B}_{kr}^{-1}X_r = \mathcal{B}_{ip}N_{pjk}X_jX_k \quad. \qquad (6.4.14)$$

Substituting (6.4.8), (6.4.12)-(6.4.14) into the second system of (6.4.5) we obtain

$$\mathcal{B}_{ip}G_{kp}\dot{X}_k = \mathcal{B}_{ip}L_{kp}X_k + \mathcal{B}_{ip}N_{pjk}X_jX_k + \mathcal{B}_{ip}F_p \quad, \qquad (6.4.15)$$

and multiplying (6.4.15) by $\mathcal{B}^{-1}$ we return to the first system of (6.4.5). This proves that both systems of (6.4.5) yield identical solutions if their initial conditions are connected via eq. (6.4.4).

To conclude, seeking other polynomial basis functions is useless, since we'll arrive to exactly the same approximate solution. On the other hand, the orthonormalization procedure can be meaningful, since it does not change the solution, but avoids inverting the Gram matrix.

## 7. Inner products with arbitrary weight

For a scalar problem, e.g. Orr-Sommerfeld or Burgers equations, the choice of the weight function in the inner product (2.18) is arbitrary. In the case of unit or Chebyshev weight, the inner products can be evaluated analytically using properties of the Chebyshev polynomials listed in Appendix A. In other cases the Gauss quadrature formulae can be efficiently applied.

An appropriate choice of the weight function can drastically improve the convergence, as was demonstrated in Gelfgat (2006) for the Burgers equation.

For the incompressible Navier-Stokes equation, use of the unit weight function allows one to eliminate the pressure by the Galerkin projection. The unity weight also yields important conservation properties of the resulting ODE system (6.1.3) and (6.1.4). At the same time, if the weight function can be optimized such that the convergence is noticeably improved, then the total number of degrees of freedom in the resulting dynamical system can be decreased. In this case it can be reasonable to give up the nice properties of the unit weight and proceed with the optimized one. Then it will be necessary to solve the pressure equation (2.11), so that the ODE system with the algebraic constraints (2.16), (2.17) will be considered. In the following we follow Gelfgat (2006) to show how the algebraic constraints can be removed in the framework of the Galerkin approach.

First of all, the Laplacian operator in pressure equation (2.11) must be supplied by boundary conditions. It was shown in Gelfgat (2006) that the boundary conditions proposed by Gresho & Sani (1987), which are limits of the momentum equation at the boundaries, yield a correct pressure field. The boundary conditions on the boundary $\Gamma$ are

$$\left[\frac{\partial p}{\partial n}\right]_\Gamma = \boldsymbol{n} \cdot \left[\frac{1}{Re}\Delta \boldsymbol{v} - (\boldsymbol{v} \cdot \nabla)\boldsymbol{v} - \frac{\partial \boldsymbol{v}}{\partial t} + \boldsymbol{f}\right]_\Gamma, \qquad (7.1)$$

Where $\boldsymbol{n}$ is the normal to $\Gamma$, and the boundary conditions for velocity are assumed to be steady. Note that Rempfer (2006) argued that the numerical solution of the pressure problem (2.11), (7.1), together with the momentum equation (2.7), do not yield a divergence-free solution for velocity. The global Galerkin method with divergence-free velocity basis functions described here does not have this problem, because the continuity equation is satisfied analytically by the basis functions, before the numerical process starts. Thus, any approximation of a solution is analytically divergence free.

For the following we represent the pressure as a truncated Chebyshev series

$$p(x,y,z,t) = \sum_{i,j,k} \vartheta_{ijk}(t) T_i(x) T_j(y) T_k(z). \qquad (7.2)$$

Here we cannot introduce any boundary conditions into the basis functions, because we cannot propose any sufficiently general change of variables that will make the boundary conditions (7.1) homogeneous for the pressure, so that they will be incorporated in the pressure basis functions. To obtain a problem for the unknown coefficients $\vartheta_{ijk}(t)$ we perform Galerkin projections of the residuals of (2.11) and (7.1) on the Chebyshev basis (7.2), and require that the projections vanish. In other words, we apply the Galerkin method separately in the domain and on the boundaries. Clearly the total number of unknowns $\vartheta_{ijk}(t)$ must be equal to the total number of equations used. Recalling that the velocity coefficients are stored in the vector $X$, we store the additional coefficients of pressure $\vartheta_{ijk}(t)$ in the vector $Y$. After the Galerkin process is completed, the system of linear algebraic equations for $Y$ has the following form

$$Q_{IJ}^{(p)} Y_J = B_{IJ}^{(p)} \dot{X}_J + L_{IJ}^{(p)} X_J + N_{IJK}^{(p)} X_J X_K + F_I^{(p)} \tag{7.3}$$

These equations are formed from the projections of Eqs. (2.11) and (7.1). The superscript $(p)$ is used to underline that all the matrices belong to the pressure problem. The matrix $Q_{ij}^{(p)}$ is singular because the pressure is defined to within an additive constant. This singularity can be easily removed by, e.g., an additional requirement $\vartheta_{000} = 0$, after which the matrix $Q^{(p)}$ is regular and its inverse is denoted as $Q^{-1}$. The Galerkin coefficients of velocity must be calculated using the equations (2.16), which take the following form

$$G_{IJ} \dot{X}_J = P_{IJ} Y_J + L_{IJ} X_J + N_{IJK} X_J X_K + F_I \ . \tag{7.4}$$

Here the first term of the r.h.s. is the projection of the pressure gradient onto the velocity basis. Substituting $Y_j$ from Eqs. (7.3) into Eqs. (7.4) we obtain

$$\tilde{G}_{IJ} \dot{X}_J = \tilde{L}_{IJ} X_J + \tilde{N}_{IJK} X_J X_K + \tilde{F}_I \ , \tag{7.5}$$

where

$$\tilde{G} = G - PQ^{-1}B^{(p)}, \quad \tilde{L} = L + PQ^{-1}L^{(p)}, \quad \tilde{N} = N + PQ^{-1}N^{(p)}, \quad \tilde{F} = F + PQ^{-1}F^{(p)} \tag{7.6}$$

Thus, after some analytical and numerical evaluations we obtain the ODE system (7.5), whose structure is equivalent to that of (6.1.1). Generally, the matrices of (7.5) do not obey the properties (6.1.3) and (6.1.4), however everything else that we have written about the system (6.1.1) is applicable also to (7.5).

Some numerical experiments comparing convergence of the Galerkin method with Chebyshev and unit weights are reported in Gelfgat (2006). There the lid-driven cavity flow and convection in a laterally cavity were taken as test problems. It was found that the Chebyshev weight allows for a better resolution of boundary layers in the convection flow, but slows down the convergence for the lid-driven flow. This was attributed to the problem of approximating the corner discontinuities.

The optimization of the weight function has never been considered for a realistic fluid dynamics problem. The only optimization example is given in Gelfgat (2006) for the Burgers equation. Considering weight functions of the form of $(x-x^2)^{-\alpha}$, $\alpha \geq 0$, it was found that the convergence is fastest at $\alpha = 1.3$.

## 8. Solved problems and other applications of the method

The effectiveness of the Galerkin approach follows from a decrease in the number of degrees of freedom of a numerical model. The decrease is usually of an order of magnitude or more. One of the earliest examples of this is shown in Fig. 3. There we consider flow in a cylinder with rotating lid. Steady states of this flow exhibit the vortex breakdown phenomenon, which was experimentally studied by Escudier (1984). At certain Reynolds number a weak reverse circulation attached to the cylinder axis appears. In tall cylinders up to three distinct recirculation zones are observed. The intensity of the reverse vortices is 3-5 orders of magnitude less than that of the main meridional circulation, which makes its numerical modeling quite challenging. We show in Gelfgat et al. 1996 that the size and position of the reverse circulations is well reproduced with 34 with 34×34 basis functions (Fig. 3a), as well as with 200×200 finite volume grid (Fig. 3c), but is resolved inaccurately with the 100×100 grid (Fig. 3b). The total number of degrees of freedom of the Galerkin method is the number of unknown Galerkin coefficients of the meridional velocity vector and that of the azimuthal velocity component, which separates as a scalar in the axisymmetric formulation. In the finite volume approach this is the number of unknown functions multiplied by the number of grid nodes.

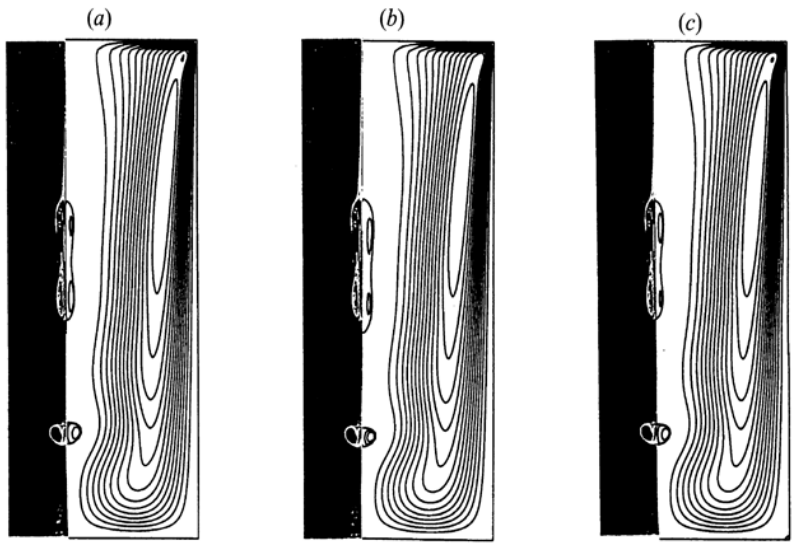

Fig. 3. Comparison of numerical results with the experimental phograph of Escudier (1984). From Gelfgat et al. (1996). The flow at H/R=3.25, Re=2752. (a) calculation with the Galerkin method using 34×34 basis functions. (b), (c) calculation with the finite volume method using 100×100 and 200×200 grids, respectively.

Clearly the total number of degrees of freedom consumed by the Galerkin method, $2 \cdot 34^2 = 2312$, is smaller than that of the finite volume method, $3 \cdot 200^2 = 120{,}000$, by about 1.5 orders of magnitude.

The next example presented in Fig. 4 is a thermocapillary convective flow in a laterally heated cavity with the aspect ratio length/height=4. The velocity boundary condition at the upper surface is

$$v = 0, \quad \frac{\partial u}{\partial y} = -Mn\frac{\partial \theta}{\partial x}, \qquad (8.1)$$

where $\theta$ is the dimensionless temperature, $Mn = Ma/Pr$, and $Ma$ and $Pr$ are the Marangoni and Prandtl numbers, respectively (other details are in Gelfgat, 2007a). The problem was treated with the truncation of 50×20 basis functions with 100 collocation points at the upper surface to satisfy the boundary condition (8.1) The critical Marangoni number, corresponding to transition from steady to oscillatory flow regime, was calculated to be $Mn_{cr} = 4781$ and the dimensionless critical frequency (imaginary part of the leading eigenvalue) $\omega_{cr} = 6674$. This result was never published because the value of $\omega_{cr}$ seemed to be too high compared with the values already known for buoyancy convection (Gelfgat et al., 1999a). Owing to the computer restrictions of that time, the convergence could not be rigorously checked. Later, the same problem was solved using a 800×200 stretched finite volume grid (Gelfgat, 2007a) and the result was $Mn_{cr} = 4779$ with the same value of the critical frequency. The convergence of the critical values with resolution obtained by the finite volume method was found to be very slow. The reason for that is seen in Fig. 4. The streamlines and the isotherms are smooth, so that the velocity and temperature fields can be easily calculated. At the same time, the perturbation patterns exhibit very steep maxima, which must be numerically resolved to obtain correct critical values. We observe here again that the Galerkin method yielded the correct result with a much smaller number of degrees of freedom.

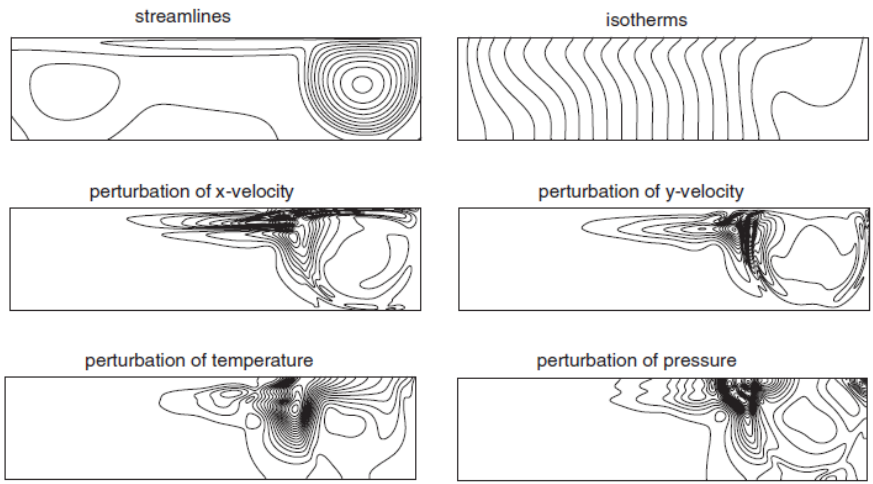

Fig. 4. Patterns of flow and amplitude of the most unstable perturbation at the critical Marangoni number for thermocapillary convection of low-Prandtl-number fluid (*Pr*=0.015) in a cavity of aspect ratio length/height=4. From Gelfgat (2007a).

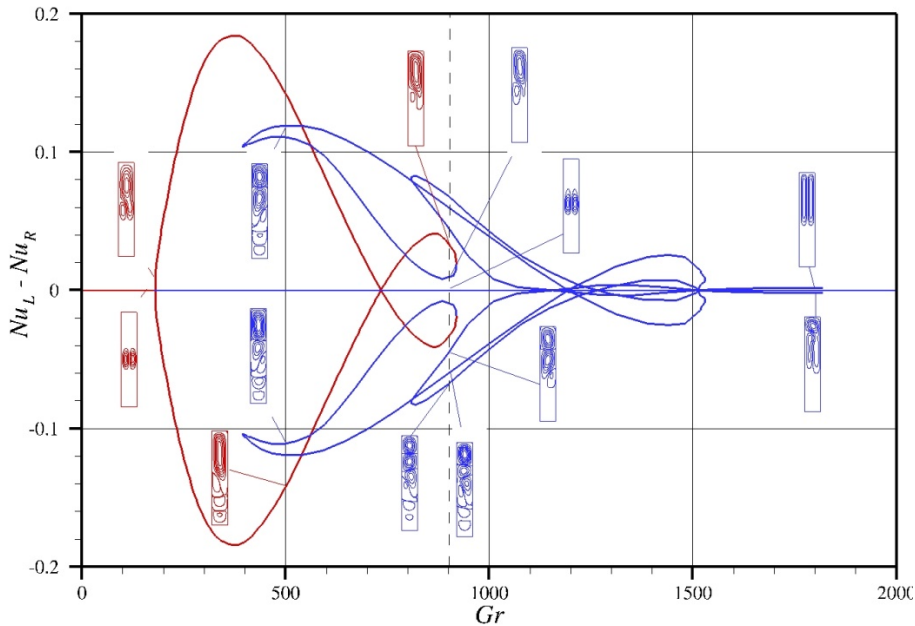

Fig. 5. Bifurcation diagram for convection in a vertical cavity with partially heated sidewall. Red lines correspond to stable steady states, blue lines to unstable ones. $Pr$=10. From Erenburg et al. (2003).

The smaller number of degrees of freedom, as well as analytic representation of the Jacobian matrix (6.1.6), allows one to effectively apply Newton iteration for calculation of the steady states. To follow different solution branches one also can apply arc-length or similar continuation technique. This is illustrated in Fig. 5 for the convection in a cavity with partially heated sidewall (Erenburg et al., 2003), the boundary conditions for which were discussed in section 4. Since the boundary conditions are symmetric, the flow at low Grashof number $Gr$ is symmetric. The symmetry is broken at $Gr = 180$. The diagram in Fig. 5 shows difference between the Nusselt numbers calculated at the left and right vertical boundaries, so that in the symmetric state the difference is zero. After the symmetry breaks, we observe several interconnected solution branches with qualitatively different flow patterns. Those depicted in red are stable, and those depicted in color are oscillatory unstable. Although most of the steady state branches are unstable, we speculate that there exist multiple oscillatory states with similar flow patterns.

The most well-known results obtained with this method are stability diagrams of swirling rotating disk – cylinder flow and buoyancy convection flows in laterally heated rectangular cavities. The first results on the three-dimensional instability of rotating disk – cylinder flow were obtained in Gelfgat et al. (2001a) using the Galerkin method with 30×30 basis functions. Since then several research groups validated these results experimentally and numerically. These comparisons are very convincing and are shown in Figs. 6 and 7. According to our results, the instability is axisymmetric for the cylinder aspect ratio (height/radius) varying between 1.6 and 2.7, and is three-dimensional outside of this interval. The three-dimensionality sets in with the azimuthal wavenumber $k = 2$ at small aspect ratios, and with $k = 3$ or 4 in taller cylinders. Several later studies tried to reproduce these results either by straightforward integration in time, or by means of stability analysis, and fully confirmed our conclusions. The quantitative comparison was carried out for the critical Reynolds numbers and critical frequencies, as well as for the azimuthal mode number. It was possible also to confirm values of the aspect ratio at which the modes replace each other.

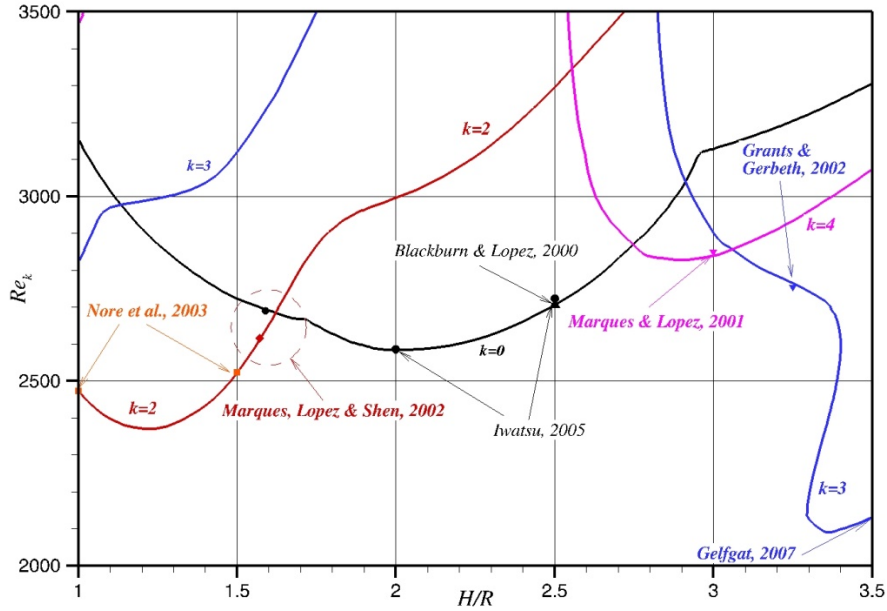

Fig. 6. Stability diagram for flow in a cylinder covered by rotating disk. The lines correspond to results obtained by the Galerkin method (Gelfgat et al., 2001) for different azimuthal wavenumbers $k$ in (3.4.1). Symbols show results independent studies.

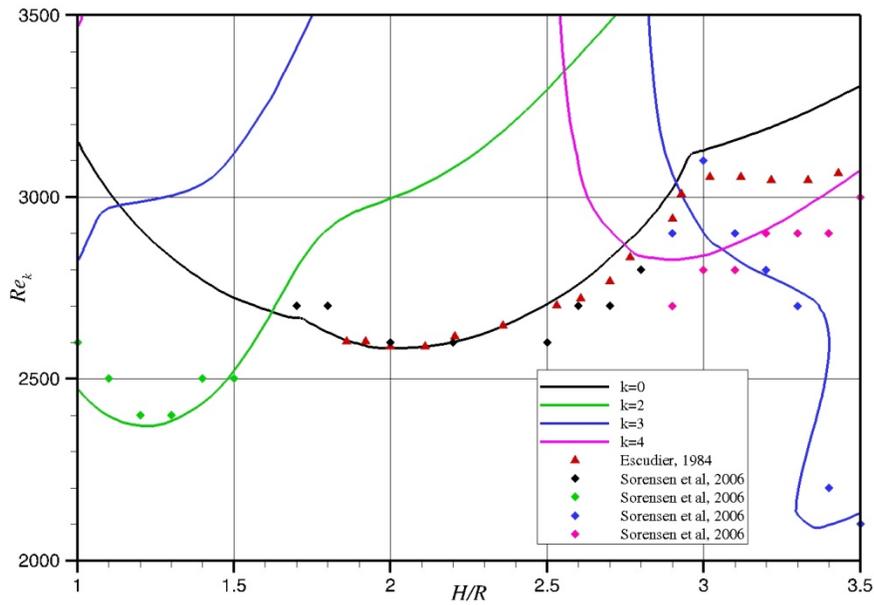

Fig. 7. Stability diagram for flow in a cylinder covered by rotating disk. The lines correspond to results obtained by the Galerkin method (Gelfgat et al., 2001) for different azimuthal wavenumbers *k* in (3.4.1). Symbols show experimental results.

A similar comparison, but experimental, was carried out by Sørensen et al (2006, 2009). Their result is shown in Fig. 7. The oscillations were measured by LDA, and the flow azimuthal periodicity by PIV. The symbols in Fig. 7 show experimentally measured points, and their color corresponds to the azimuthal wavenumber as shown in the figure. The pioneer results of Escudier (1984) are also shown. Taking into account experimental uncertainties, the agreement between experiment of Sørensen et al (2006) and the numerical predictions of the Galerkin method is quite impressive. The agreement with earlier results of Escudier (1984) is observed only for the axisymmetric mode, but his experiments were based only on visualizations by the aluminum powder, and the instability was observed only as oscillations of the vortex breakdown bubbles.

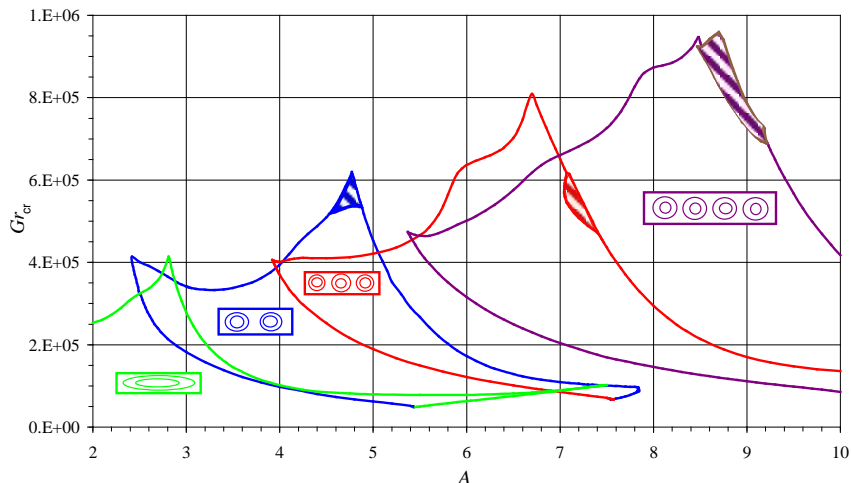

Fig. 8. Stability diagram for convective flow in laterally heated cavities. The curves color corresponds to the number of convective rolls in the flow pattern. The flows are stable below or between the curves of the same color. Dashed regions correspond to the stability regions of similar flows with broken rotational symmetry. *Pr*=0. From Gelfgat et al. (1999)

Other stability

results obtained by the described Galerkin method for similar swirling flows can be found in Gelfgat at al. (1996b) for flow in a cylinder with independently rotating top and bottom, and in Marques et al. (2003) for independently rotating top and sidewall.

The neutral curves shown in Fig. 6 are plotted through up to a hundred calculated critical points. The next example, relating to the oscillatory instability of buoyancy convection flows in laterally heated cavities and shown in Fig. 8, required several hundreds of critical points to complete the study. The calculations were performed with up to 60×20 basis functions. With increasing cavity aspect ratio $A$= length/height, and at large enough Grashof number, the single convective cell splits into several cells. The number of cells grows with the aspect ratio. At the same time several steady states with a different number of rolls can be stable at the same set of governing parameters, as shown in Fig. 9. The transition from one number of cells to another takes place at points where a neutral curve of a certain color continues with a different color. At these points the flows with, e.g., two and three rolls are indistinguishable. Other results on stability of convection in rectangular cavities can be found in Gelfgat et al. (1996, 1999a,b), Erenburg et al. (2003), and Gelfgat (2004). One particularly interesting result reported in Gelfgat (2004) showed that weakly non-linear approximation of limit cycles (6.9)-(6.11) yields results that are very close to those obtained by independent straightforward time integration.

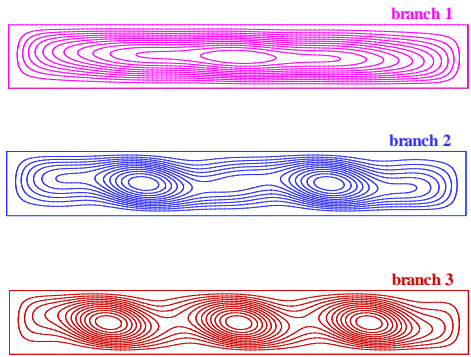

Fig. 9. Three distinct stable steady states found at $Pr$=0, A=7, Gr=88,000. From Gelfgat & Bar-Yoseph (2004).

Several studies have been devoted to three-dimensional instabilities of axisymmetric buoyancy convection in vertical cylindrical containers. These studies were started in Gelfgat et al. (1999c), where we were able to reproduce a nice experimental result showing the breaking of axisymmetry leading to a spoke pattern with azimuthal wavenumber $k = 9$. Later we studied cylinders with non-uniformly heated sidewalls that mimicked conditions of Bridgman crystal growth (Gelfgat et al., 2000, 2001b). Later works were devoted to axisymmetric flows driven by rotating or traveling magnetic field (Gelfgat & Gelfgat, 2004; Gelfgat, 2005). Most of these results were reviewed in more detail in Gelfgat & Bar-Yoseph (2004).

Additional opportunities are provided by analytical representation of numerical solution via the Galerkin series. Clearly, one can differentiate or integrate the series without any noticeable loss of accuracy, contrary to low-order methods. An obvious example is the calculation of flow trajectories using a previously calculated steady or time-dependent flow. Since the velocity field is defined analytically in the whole domain, wherever the liquid particle arrives, its velocity is known without the need to interpolate between grid nodes. This fact was used in Gelfgat (2002), where trajectories were calculated over very long time to obtain a Poincare map in the midplane.

Another application of the divergence-free bases (3.3.2), (3.3.3), and (3.3.6) is visualization of three-dimensional incompressible flows, as described in Gelfgat (2014). Without going into detail, we only mention that projection of flow on each divergence-free set can be interpreted as a divergence-free projection on coordinate planes $x = const, y = const, z = const$, which results in two-component divergence-free fields. These can be described by an analog of the streamfunction. Assembling all the planes, e.g., $x = const$, we obtain a scalar 3D function whose isosurfaces are tangent to the projected vectors. Three such projections of three sets of coordinate planes complete the visualization of a three-dimensional flow field. Details and illustrations can be found in Gelfgat (2014).

## 9. Similar approaches in studies of other authors

As mentioned, the idea to use linear superpositions of Chebyshev polynomials for definition of basis functions satisfying linear and homogeneous boundary conditions was introduced by Orszag (1971a,b) for the homogeneous two-point Dirichlet problem. Since then, similar linear-superposition-based basis functions were used to solve one-dimensional problems for, e.g., Orr-Sommerfeld and boundary layer equations, by Holte (1983), Pasquarelli (1991), Yueh & Weng (1996), Yang (1997), Borget et al. (2001), Yahata (2001), Bistrian et al. (2009), and Buffat & Le Penven (2011). In these works the basis functions were constructed from the Chebyshev polynomials. Recently, Wan & Yu (2017) applied the same idea to Legendre polynomials. Grants & Gerbeth (2001), Uhlmann & Nagata (2006), and Batina et al. (2009) used the same approach for a two-dimensional flow field, but their basis functions were not divergence-free. Picardo et al. used linear superpositions for a two-fluid problem, as was done in Gelfgat et al. (2001d).

Moser et al. (1983) proposed to multiply linear superpositions of the Chebyshev polynomials by powers of the Chebyshev weight function in order to make better use of the polynomial orthogonality properties. For problems with two periodic directions, these authors constructed a divergence free basis, in which the non-periodic direction was treated by linear superpositions of the Chebyshev polynomials multiplied by additional weight-dependent functions. Such functions were used for either coordinate or projection bases in the weighted residuals method by Ganske et al. (1994), Goddeferd & Lollini (1999), Kerr (1996). It should be noted that multiplication by either function makes evaluation of derivatives and computations of their Galerkin projections more complicated.

Yahata (1999) solved a problem of buoyancy convection in laterally heated cavities similar to those treated by Gelfgat & Tanasawa (1994) and later by Gelfgat et al. (1997 1999a,b). He used linear superpositions of Chebyshev polynomials to construct basis functions for the temperature and the streamfunction with subsequent orthogonalization of the basis. The inner product was formulated with an arbitrary weight function, however it is not clear which weight was used. By evaluating derivatives of the streamfunction basis of Yahata (1999) one would obtain the two-dimensional basis (3.2.3), so that both formulations are equivalent in the 2D case. An extension of Yahata's approach to 3D formulation would require replacement of the stream function by a vector potential, which would complicate the formulation.

Suslov & Paolucci (2002) also solved similar convection problem in a cavity with coordinate functions (3.2.3). For the projection system they used the same functions multiplied by the Chebyshev weight, which made the Gram matrix sparser and allowed to use the fast Fourier transform (FFT) to evaluate non-linear terms of the dynamical system. The whole approach was used for straightforward integration in time, but did not exhibit much advantage compared to other methods and, to the best of the author's knowledge, had no further continuation.

**10. To conclude: what else can be done?**

To discuss further possible implementations of this Galerkin approach we mention that with the present growth of computational power and state-of-the-art methods of numerical linear algebra, the solution of two-dimensional and quasi-two-dimensional problems became feasible, and sometimes more efficient, with lower order methods. A very popular methodology of turning a time-stepping code into a steady state / stability solver can be found in Boronska & Tuckerman (2010a,b) and Tuckerman et al (2018). Another possible methodology together with several examples are given in Gelfgat (2007a,b). For these problems the Galerkin approach can be more suitable for weakly non-linear bifurcations analysis. It is not clear, however, whether results applicable only for small supercriticality will justify the effort.

One of possible applications of the method is consideration of fully three-dimensional flows, steady and unsteady, in axisymmetric domains. In these problems the bases (3.4.5) and (3.4.6) can be combined with the Fourier decomposition in the circumferential direction, so that the Gram matrices will separate for each Fourier mode and thus not be too large, so that they will be easily inverted. Finally, one obtains an ODE system, where equations corresponding to different Fourier modes will be coupled via the non-linear terms. This system allows for computation of steady states, path-following, stability analysis and time-dependent calculations (see, e.g., Boronska & Tuckerman, 2010a,b).

Possibly, the most challenging task would be to develop a fully three-dimensional solver for flow in a three-dimensional rectangular box. This would require orthogonalization of the whole set of bases (3.3.2), (3.3.3), and (3.3.6) with subsequent effective treatment of the non-linear terms. If successful, this approach would allow one to have steady state, stability, and time-dependent solvers within a single computational model, with which very complicated flows can be studied. It should be mentioned here that Krylov subspace based solvers are sometimes thought to be applicable only to sparse matrices, for which matrix-vector products can be quickly evaluated. However, matrix-vector products are also fast for the method that we have described, even though all the related matrices are densely filled.

As mentioned in the very beginning of this paper, the Galerkin approach that we have described is limited to simple domains, which must be curvilinear rectangles, in other words, regions bounded by coordinate surfaces. This is a very stringent restriction since it excludes a very big set of important problems, in which the boundaries have a more complicated shape. Another restriction for implementation of this method is flows with deformable interfaces. One of the ways to solve such problems on fixed grids is the immersed boundary method and/or the diffuse interface approach (not described here). Implementation of these approaches for the spectral method would require a good approximation of the delta function, which can be difficult to do using smooth polynomials.

Finally, we emphasize a technique for flow visualization made in Gelfgat (2014). This is applicable to flows calculated by any of numerical method, and can be very helpful for understanding the topology of complicated three-dimensional flows.

## Appendix A. Shifted Chebyshev polynomials and some of their useful properties

Shifted Chebyshev polynomials of the first and the second type shifted onto the interval $0 \leq x \leq 1$ are defined as

$$T_n(x) = \cos[n \arccos(2x-1)], \quad U_n(x) = \frac{\sin[(n+1)\arccos(2x-1)]}{\sin[\arccos(2x-1)]}, \quad 0 \leq x \leq 1 \quad (A1)$$

The two systems $\{T_n(x)\}_{n=1}^{\infty}$ and $\{U_n(x)\}_{n=1}^{\infty}$ form bases in $L_2[0,1]$ and are connected via the relation

$$T_n'(x) = 2(n+1)U_{n-1}(x), \quad (A2)$$

which resembles connection between sine and cosine. Values of the polynomials and their derivatives in the points $x = 0$ and $x = 1$ that are needed to define basis functions for different boundary conditions are

$$T_n(0) = (-1)^n, \qquad\qquad T_n(1) = 1 \quad (A3)$$

$$U_n(0) = (-1)^n(n+1), \qquad\qquad U_n(1) = n+1 \quad (A4)$$

$$T_n'(0) = (-1)^n 2n^2, \qquad\qquad T_n'(1) = 2n^2 \quad (A5)$$

$$U_n'(0) = (-1)^n n(n+1)(n+2)/3, \qquad\qquad U_n'(1) = n(n+1)(n+2)/3 \quad (A6)$$

For the following we assume that for $k > 0$, $T_{-k}(x) = T_k(x)$ and $U_{-k}(x) = U_{-k-1}(x) = U_k(x)$. To evaluate inner products we need to decompose the polynomial derivatives and the polynomials products into polynomial sums. For the multiplication of a polynomial by a polynomial we have

$$T_n(x)T_m(x) = \frac{1}{2}(T_{n+m}(x) + T_{n-m}(x)) \quad (A7)$$

$$T_n(x)U_m(x) = \frac{1}{2}(U_{n+m}(x) + U_{n-m}(x)) \quad (A8)$$

$$U_n(x)U_m(x) = \sum_{k=0}^{n} U_{m-n+2k}(x) \quad (A9)$$

The derivatives can be represented as Chebyshev series as

$$T_{n+l}^{l+1}(x) = 2^{l+2} l! (n+l) \sum_{j=0}^{[(n-1)/2]} a_{n-1-2j} \binom{j+l}{l}\binom{n-j+l-1}{l} T_{n-1-2j}(x) \quad (A10)$$

$$U_{n+l}^{l+1}(x) = 2^{l+3}(l+1)! \sum_{j=0}^{[(n-1)/2]} a_{n-1-2j} \binom{j+l+1}{l+1}\binom{n-j+l}{l+1} T_{n-1-2j}(x) \quad (A11)$$

$$a_0 = \frac{1}{2}, \quad a_{m>0} = 1$$

For example,

$$T'_n(x) = 4n \sum_{j=0}^{[(n-1)/2]} a_{n-1-2j} T_{n-1-2j}(x) \tag{A12}$$

Here $\binom{m}{n}$ is the binomial coefficient. After the basis functions are built, the Galerkin projections can be computed with the unity weight

$$\langle f, g \rangle_1 = \int_0^1 f(x)g(x)dx, \tag{A13}$$

or with the Chebyshev weight

$$\langle f, g \rangle_{Ch} = \int_0^1 (x - x^2)^{-1/2} f(x)g(x)dx, \tag{A14}$$

or with an arbitrary weight. For example, in Gelfgat (2005) we used

$$\langle f, g \rangle_a = \int_0^1 (x - x^2)^{-\alpha} f(x)g(x)dx, \qquad 0 < \alpha < 1 \tag{A15}$$

All the inner products needed to complete the Galerkin procedure can be evaluated analytically if either the unity or Chebyshev weight is implied. The following relations (base products) are needed for that

$$\langle T_n(x), T_m(x) \rangle_1 = \frac{1}{8}[1 + (-1)^{n+m-1}]\left[\frac{1}{n+m+1} - \frac{1}{n+m-1} + \frac{1}{n-m+1} - \frac{1}{n-m-1}\right] \tag{A16}$$

$$\langle T_n(x), T_m(x) \rangle_{Ch} = a_n \pi \delta_{nm} \tag{A17}$$

And $\delta_{nm}$ is the Kronecker symbol. The relation (A2), (A7)-(A16) allow one to reduce all the inner products to the above ones. In the case of arbitrary inner product the base products $\langle T_n(x), T_m(x) \rangle$ must be evaluated numerically, which is usually done using the Gauss quadrature. Then all the other products can be expressed as sums using relations (A2) and (A7)-(A11). Additionally, for evaluation of inner products in orthogonal curvilinear coordinates, one may need the following relation

$$x^m = 2^{-2m+1} \sum_{i=0}^{m} a_{m-i} \binom{2m}{i} T_{m-i}(x) \tag{A18}$$

Finally, to calculate the shifted Chebyshev polynomials in a point, the following recurrent formulae can be used

$$T_n(x) = (4x - 2)T_{n-1}(x) - T_{n-2}(x) \tag{A19}$$

$$U_n(x) = (4x - 2)U_{n-1}(x) - U_{n-2}(x) \tag{A20}$$

Further details can be found in the books of Paszkowski (1975) and Mason & Handscomb (2003).